\PassOptionsToPackage{svgnames,table}{xcolor}
\documentclass[11pt,reqno]{article}
\usepackage{amsmath,hyperref,amssymb, amsthm}
\usepackage{authblk}
\usepackage[all]{xy}
\usepackage{graphicx,url}
\usepackage{color}
\usepackage{slashed}
\usepackage{tikz}
\usepackage{lipsum}
\usepackage{caption}
\usepackage{ytableau}
\usepackage{longtable}
\usepackage{multirow}
\usepackage[table]{xcolor}
\usepackage{array}
\usepackage{booktabs}

\usepackage{caption}

\usepackage[style=ieee,citestyle=numeric-comp]{biblatex}
\newcommand{\be}{\begin{eqnarray}}
\newcommand{\ee}{\end{eqnarray}}
\newcommand{\eeq}{\end{equation}}
\newcommand{\beq}{\begin{equation}}

\allowdisplaybreaks \numberwithin{equation}{section}
\renewcommand{\Large}{\large} 

\DeclareSymbolFont{AMSa}{U}{msa}{m}{n}
\DeclareSymbolFont{AMSb}{U}{msb}{m}{n}
\DeclareMathSymbol{\fieldR}{\mathalpha}{AMSb}{"52}

\renewcommand{\Im}{\imag}
\DeclareMathOperator{\imag}{Im}

\newcommand{\CA}{{\cal A}}
\newcommand{\CB}{{\cal B}}

\newcommand{\CH}{\mathcal{H}}
\newcommand{\Hh}{\mathcal{H}}

\newcommand{\CI}{{\cal I}}

\newcommand{\CL}{{\cal L}}
\newcommand{\CN}{\mathcal{N}}
\newcommand{\CM}{\mathcal{M}}

\newcommand{\calC}{\mathcal{C}}

\newcommand{\TDL}{\mathrm{TDL}}

\newcommand{\NN}{\mathbb{N}}
\newcommand{\ZZ}{\mathbb{Z}}
\newcommand{\RR}{\mathbb{R}}
\newcommand{\CC}{\mathbb{C}}
\newcommand{\QQ}{\mathbb{Q}}

\DeclareMathOperator{\End}{End}
\newcommand{\EndL}{\mathsf{L}}

\def\beq{\begin{equation}}
\def\eeq{\end{equation}}
\def\bea{\begin{eqnarray}}
\def\eea{\end{eqnarray}}

\def\<{\langle}

\newcommand{\rrangle}{\rangle\!\rangle}
\newcommand{\llangle}{\langle\!\langle}

\usepackage[most]{tcolorbox}
\usepackage{amsthm}

\newcounter{definition}
\renewcommand{\thedefinition}{\arabic{definition}}

\newtcolorbox{definition}[1][]{
    enhanced,
    breakable,
    colback=gray!10,
    colframe=gray!60,
    boxrule=0.8pt,
    sharp corners,
    fontupper=\itshape,
    fonttitle=\bfseries,
    coltitle=black,
    attach boxed title to top left={
        xshift=5mm,
        yshift=-2mm,
    },
    boxed title style={
        sharp corners,
        colback=blue!15,
        colframe=blue!50,
        boxrule=0.8pt,
    },
    code={\refstepcounter{definition}},
    title={Definition~\thedefinition},
    #1
}

\usepackage{tcolorbox}
\tcbuselibrary{skins,breakable}
\usetikzlibrary{shadings,shadows}
    {\endtcolorbox}

    {\endtcolorbox}

    {\endtcolorbox}

\title{Hodge Loci and Complex Multiplication via\\ Generalized Symmetries in Calabi-Yau sigma models}
\author[1]{Roberta Angius\thanks{roberta.angius@uni-hamburg.de}}
\author[2,3]{Roberto Volpato\thanks{volpato@pd.infn.it}}
{\small \affil[1]{\small  II. Institut f\"ur Theoretische Physik, Universit\"at Hamburg, Notkestrasse 9, 22607 Hamburg, Germany}
\affil[2]{\small Dipartimento di Fisica e Astronomia, Universit\`a di Padova, Via Marzolo 8, 35131, Padova, Italy}
\affil[3]{\small INFN, sez. di Padova, Via Marzolo 8, 35131, Padova, Italy}}

\begin{document}

\maketitle

\abstract{We propose a sigma-model analogue of Hodge loci in the moduli space of geometric Calabi-Yau compactifications, characterized by the emergence of non-trivial rational Hodge endomorphisms, using generalized symmetries. In the CFT description, the complex cohomology is spanned by Ramond-Ramond ground states, the Hodge decomposition is determined by the $U(1)\times U(1)$ R-charges, and the rational structure is provided by BPS boundary states, with polarization induced by the open string Witten index. Hodge loci are identified by the existence of a non-trivial category $\TDL$ of topological defects preserving the $\mathcal{N}=(2,2)$ superconformal algebra and acting invertibly on the spectral-flow generators. At special points on these loci, the category $\TDL$ exhibits additional arithmetic structure and admits embeddings of finite products of number fields with Complex Multiplication, leading to stronger constraints on the boundary states of the theory. Although the construction is general, we analyze in detail the cases of elliptic curves and $K3$ surfaces.}

\newpage

\tableofcontents
\newpage

\section{Introduction}

    String theory compactifications on Calabi–Yau manifolds have long provided one of the most promising frameworks for exploring lower-dimensional phenomenology. Despite the extensive progress, one of the most central and still open questions in the community remains how to achieve a deeper and more complete understanding of the physics associated with the various geometries encountered throughout the corresponding moduli spaces.  From the point of view of the effective field theories associated with these compactifications, the period vectors of the compact geometry constitute one of the fundamental ingredients, as they determine the exact form of the low-energy effective action through quantities such as the prepotential, the kinetic functions, and the Yukawa couplings \cite{Ceresole:1994cdf,Craps:1997cbr}. A detailed understanding of the structure of these periods should therefore provide a major advance in the study of low-energy phenomenology and allow one to test many Swampland conjectures in these broad classes of compactifications \cite{Vafa:2005swamp} (see \cite{Brennan:2018bcv, Palti:2019,Van_Beest:2022vcv} for reviews). It is well known that period vectors can be computed through Picard–Fuchs equations \cite{Hosono:1995hkt, Hosono:1995kty}, whose solutions are in general complicated functions of the moduli. A great effort has been devoted in both the mathematical and physical literature to understanding the functional form of these objects. However, apart from a limited number of explicit examples (see \cite{CANDELAS:199121} for a prime example), a general understanding is still lacking, except in some special regions of moduli space, such as asymptotic regions near singularities, where the machinery of Hodge theory allows one to approximate the periods by polynomial expressions, supplemented by infinite series of exponentially suppressed corrections (see \cite{Cecotti:2020, Grimm:2018ohb, Grimm:2018cpv, Grimm:2019bey, Bastian:2020egp, Bastian:2021eom} for the analytic structure of the periods and \cite{Lee:2022tim, Alvarez_Garcia:2024, Alvarez_Garcia:2025, Hassfeld:2025uoy, Monnee:2025ham} for the corresponding underlying geometry) that become increasingly relevant as one moves toward the bulk of moduli space, which remains largely unexplored. \\
    Another way to approach string theory compactifications on Calabi–Yau manifolds is through the corresponding two-dimensional worldsheet theories with target space $\mathcal{M}_{1,9-2d} \times X_d$, where $X_d$ is a compact Calabi–Yau manifold, or a complex torus, of complex dimension $d$. For such geometries, the internal sector of the worldsheet superconformal field theory (SCFT) is described by a non-linear sigma model (NLSM) with $\mathcal{N}=(2,2)$ supersymmetry. In this paper, we adopt this perspective and focus on the moduli space of these $\mathcal{N}=(2,2)$ NLSMs, which already encodes the information needed to compute physical couplings while also allowing for extensions of the setup to non-geometric backgrounds, such as asymmetric orbifolds.
    Nevertheless, even within the moduli space of these $\mathcal{N}=(2,2)$ NLSMs, only special points corresponds to exactly solvable models: these include rational conformal field theories (RCFTs) and toroidal orbifold compactifications.
    
    Recently, new light has been shed on certain internal regions of the geometric moduli space corresponding to loci where new discrete symmetries, acting on the moduli, are realized \cite{grimm:2024tgd}. In mathematical terms, these correspond to special cases of Hodge loci, where non-trivial Hodge tensors, absent at generic points of the moduli space, appear. The emergence of such tensors, implementing $\mathbb{Z}$-linear transformations on the integral (or rational) middle cohomology which complex extension preserves the Hodge decomposition, imposes algebraic constraints on the periods which lead to simplification of the corresponding towers of exponential corrections. In the present work we propose a sigma-model analogue of these Hodge loci, with the advantage that, from the CFT perspective, a broader class of symmetries becomes available, including non-geometric and non-invertible symmetries.

    The central objects in our proposal, playing the role of integral/rational Hodge tensors, are Topological Defect Lines (TDLs) implementing invertible or non-invertible (generalized) symmetries of the underlying $2d$ SCFT. Over the past two decades, a considerable amount of work has been devoted to characterizing these objects in various classes of two-dimensional conformal field theories \cite{Verlinde:1988sn, Petkova:2000ip,Frohlich:2004ef, Frohlich:2009gb, Frohlich:2006ch, Chang_2019, Carqueville:2023jhb, Bhardwaj:2017xup, Fuchs:2007tx, Bachas:2012bj, Thorngren:2019rtw, Thorngren:2021rtw, Angius:2024evd, Volpato:2024goy, Angius:2025zlm, Angius:2025dcu, Angius:2025ium}. It is known that the natural mathematical framework for describing these objects is that of tensor categories, whose objects are precisely topological defects equipped with a (generically non-invertible) fusion structure. Classical examples include Tambara–Yamagami categories, such as the one realized in the Ising model through three simple defects, two of which implement the $\ZZ_2$ symmetry, while the third implements the Kramers–Wannier duality, or again Verlinde lines \cite{Verlinde:1988sn} preserving the full chiral algebra in rational conformal field theories (RCFTs). In boundary conformal field theories, boundary states furnish modules over these categories, providing an additional framework to study their properties and, at the same time, acquiring a physically interesting interpretation in the context of Calabi–Yau sigma models as operators acting on D-branes. This interpretation was first exploited in \cite{Angius:2024evd} to establish general properties of a special class of topological defects, namely those preserving the full $\CN=(4,4)$ superconformal algebra together with the spectral flow operators, in NLSMs with $K3$ target space. In the present work, we consider a generalization of this class of defects by relaxing the conditions on the preserved symmetry algebra and on the action on the spectral flow operators, while extending the framework beyond NLSMs with $K3$ target space.
    
\begin{figure}[h!]
        \centering
        \includegraphics[width=0.64\linewidth]{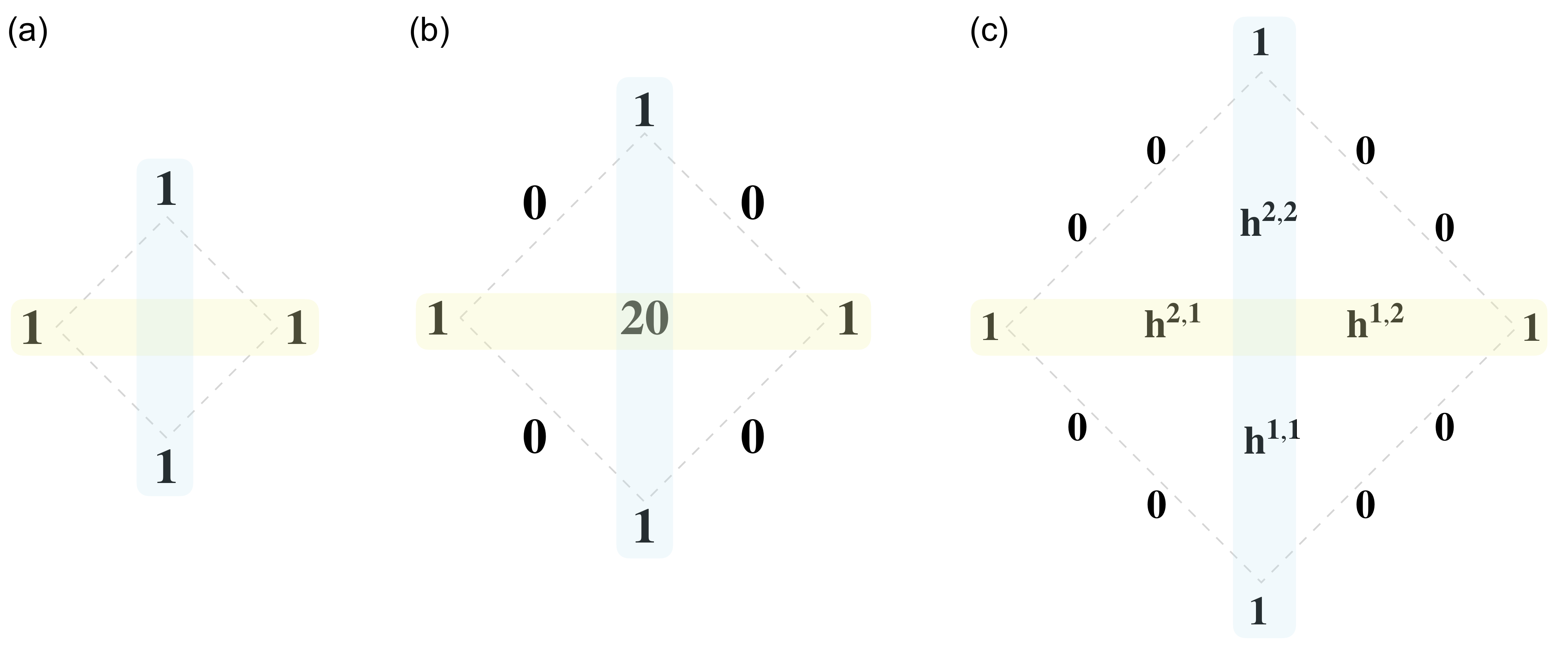}
        \caption{Hodge diamonds of elliptic curves (a), $K3$ surfaces (b) and Calabi-Yau three-folds (c).}
        \label{fig:Hodge_diamond}
    \end{figure}

    The framework of our proposal consists of (Euclidean, compact, unitary) $2d$ SCFTs $\mathcal{C}$ with $\mathcal{N}=(2,2)$ superconformal algebra and central charges $c= \bar{c}=3d$, with $d \in \mathbb{N}$, realized for example by NLSMs with target space a Calabi–Yau manifold $X_d$ of complex dimension $d$ (including complex tori).  The starting point is the Hilbert space sector $\mathcal{H}(\mathcal{C},\CC)$ of Ramond–Ramond ground states of the theory, which can be identified, as a vector space, with the complex cohomology of the target geometry, $\mathcal{H}(\mathcal{C},\CC) \cong H^{\bullet}\left(X_d, \mathbb{C} \right)$.  The $\mathcal{N}=(2,2)$ superconformal algebra provides two $U(1)$ R-symmetries associated with the holomorphic and anti-holomorphic sectors. The charges $q, \tilde{q}$ of the representations under these two symmetries induce a grading that equips the vector space $\mathcal{H}(\mathcal{C},\CC)$ with a Hodge decomposition:
    \begin{equation}
        \CH(\calC,\CC) =\bigoplus_{r,s=0}^d \CH^{r,s}(\calC,\CC).
        \label{e:intro_Hodge_decomp}
    \end{equation}
    The next ingredient is provided by the BPS boundary states $\vert \vert \mathcal{B} \rrangle$ of the theory, whose charges under the massless R–R fields define a charge vector $\Gamma_{\mathcal{B}}$ with $\dim \mathcal{H}(\mathcal{C},\CC)$ components. The set of all $\mathbb{Z}$-linear combinations of charges vectors $\Gamma_{\mathcal{B}}$ associated with BPS boundary states spans a lattice $\mathcal{H}_{\bullet} (\mathcal{C}, \mathbb{Z})$, equipped with the bilinear pairing induced by the open-string Witten index.  Its dual lattice defines an integral subset of the cohomology $\mathcal{H}^{\bullet} (\mathcal{C}, \mathbb{Z}) \subset \mathcal{H}^{\bullet} (\mathcal{C}, \mathbb{C})$. Due to the boundary gluing conditions, BPS boundary states can only carry charges under R–R ground states satisfying either $q=\tilde{q}$, spanning the middle part of the cohomology $\mathcal{H}^{middle} (\mathcal{C}, \mathbb{C}) \subset \mathcal{H}^{\bullet} (\mathcal{C}, \mathbb{C})$, corresponding to the so-called A-type branes, or $q=-\tilde{q}$, spanning the vertical part $\mathcal{H}^{vert} (\mathcal{C}, \mathbb{C}) \subset \mathcal{H}^{\bullet} (\mathcal{C}, \mathbb{C})$, corresponding to the so-called B-type branes. The rational extension of the lattice $\mathcal{H}^{\bullet} (\mathcal{C}, \mathbb{Q}) = \mathcal{H}^{\bullet} (\mathcal{C}, \mathbb{Z}) \otimes \mathbb{Q}$ allows to define a rational Hodge structure $\left( \mathcal{H}^{\bullet} (\mathcal{C}, \mathbb{Q}), \mathcal{H}^{\bullet} (\mathcal{C}, \mathbb{C}) \right)$ with
    \begin{equation}
        \mathcal{H}^{\bullet} (\mathcal{C}, \mathbb{C}) = \mathcal{H}^{\bullet} (\mathcal{C}, \mathbb{Q}) \otimes \mathbb{C}
    \end{equation}
   preserving the Hodge grading and equipped with a polarization induced by the complex extension of the Witten pairing. Notice that for elliptic curves, $K3$ surfaces and $CY_3$, the middle and vertical cohomologies together span the entire Hodge diamond (see Figure \ref{fig:Hodge_diamond}). This approach is in the same spirit as \cite{Jockers:2025fgv}, but our focus is primarily on topological defects and it involves models with guaranteed spacetime supersymmetry.

    Equipped $\mathcal{C}$ with a rational Hodge structure $\left( \mathcal{H}^{\bullet} (\mathcal{C}, \mathbb{Q}), \mathcal{H}^{\bullet} (\mathcal{C}, \mathbb{C}) \right)$, the next step consists in identifying the objects in the theory that play the role of Hodge endomorphisms (Hodge tensors), namely endomorphisms of the $\mathbb{Q}$-vector space $\mathcal{H}^{\bullet} (\mathcal{C}, \mathbb{Q})$ that extend by $\mathbb{C}$-linearity to transformations of $\mathcal{H}^{\bullet} (\mathcal{C}, \mathbb{C})$ preserving the Hodge decomposition \eqref{e:intro_Hodge_decomp}.  As anticipated above, we identify these objects with Topological Defects Lines $\mathcal{L}$ of $\mathcal{C}$ that $(1)$ preserve the $\mathcal{N}=(2,2)$ superconformal algebra and $(2)$ act invertibly on the spectral flow operators. The collection of such defects forms a category, which we denote by $\TDL$. The motivation for imposing these two requirements originates from the underlying string theory compactified on $\mathcal{M}_{1,9-2d} \times X_d$: we seek topological defects of the full worldsheet CFT that act only on the internal sector $\mathcal{C}$ while preserving both the physical-state condition for fundamental strings (property $(1)$) and the spacetime supersymmetry interpretation (property $(2)$). In particular, each defect $\mathcal{L} \in \TDL$ induces an endomorphism $L= \hat{\mathcal{L}}_{\vert \mathcal{H}^{\bullet}(\mathcal{C}, \ZZ)}$, which maps BPS D-branes, represented by boundary states $\vert \vert \mathcal{B} \rrangle$ with associated charge vectors $\Gamma_{\mathcal{B}}$, into BPS D-branes $\hat{\mathcal{L}} \vert \vert \mathcal{B} \rrangle$, with charge vector $\Gamma_{\mathcal{LB}}$, remaining within the same string model. We define Hodge loci to be those points in the moduli space of $\mathcal{N}=(2,2)$ 
    SCFTs
    at which there exists at least one non-trivial defect $\mathcal{L} \in \TDL$ that is absent at generic points. This statement constitutes our first main Definition \ref{d:Hodge_loci} in Section \ref{s:proposal}. The dictionary of the proposal, including the comparison with the geometric Hodge theory analogue, is summarized in Table \ref{t:dictionary}.

\begin{table}[h!]
\centering

\renewcommand{\arraystretch}{1.25}
\setlength{\arrayrulewidth}{0.6pt}
\arrayrulecolor{black} 

\begin{tabular}{|>{\centering\arraybackslash}m{6.3cm}|
                >{\centering\arraybackslash}m{7.4cm}|}

\hline

\rowcolor{blue!15}
\rule{0pt}{3.2ex}
\Large\textbf{\textsc{Hodge Theory}} &
\Large\textbf{\textsc{CFT}} \\
\hline

\rowcolor{gray!3}
\rule{0pt}{3.2ex}
$CY_d=$ Calabi--Yau geometry &
$\mathcal{C}=\{\mathcal{N}=(2,2)\ \mathrm{SCFT}\ \text{with target }CY_d\}$ \\
\hline

\rowcolor{gray!3}
Complex cohomology $H^{\bullet}(CY_d,\mathbb{C})$ &
Hilbert space of R--R ground fields $\mathcal{H}^{\bullet}(\mathcal{C}, \mathbb{C})$ \\
\hline

\rowcolor{gray!3}
Hodge decomposition &
$U(1)\times U(1)$ R-charge eigenspaces \\
\hline

\rowcolor{gray!3}
Integral Hodge structure &
R--R D-brane charges \\
\hline

\rowcolor{gray!3}
Polarization &
Open string Witten index \\
\hline

\rowcolor{gray!3}
Hodge endomorphisms &
Generalized symmetries \\
\hline

\rowcolor{gray!3}
Algebra of endomorphisms &
Category $\mathrm{TDL}$ of topological defects \\
\hline

\end{tabular}

\caption{Dictionary of the proposal showing the analogous counterparts in the geometric setting.}
\label{t:dictionary}

\end{table}

The second main outcome of the proposal is the definition of Complex Multiplication (CM) type in the CFT setting (Definition \ref{d:CM_cft} in Section \ref{s:proposal}). Roughly speaking, this occurs whenever there exists a subcategory $\TDL' \subseteq \TDL$ whose associated algebra decomposes as a finite product of CM fields, $K_1 \times ... \times K_n$, with one factor associated with each irreducible component of the rational Hodge structure $\left( \mathcal{H}^{\bullet} (\mathcal{C}, \mathbb{Q}), \mathcal{H}^{\bullet} (\mathcal{C}, \mathbb{C}) \right)$. This condition is highly restrictive and imposes strong constraints on the BPS boundary states of the theory. The key property is that, at these points, each irreducible component of  $\left( \mathcal{H}^{\bullet} (\mathcal{C}, \mathbb{Q}), \mathcal{H}^{\bullet} (\mathcal{C}, \mathbb{C}) \right)$, which may be interpreted as a set of BPS boundary states, is $1$-dimensional as a vector space over the corresponding number field $K_i$. Consequently, once $K_i$ and a single boundary state in the associated irreducible component are known, the entire set of boundary states within that component can be generated.

Returning to the comparison with the geometric case, the Hodge loci in the CFT setting, namely the loci where new generalized symmetries of the type described above appear, can be viewed as loci where the periods satisfy additional algebraic constraints. In this context, the periods are the entries of the R-R charge vectors $\Gamma_{\mathcal{B}}$, which may be understood as the integrals of the R-R forms spanning $\mathcal{H}^{\bullet}(\mathcal{C},\CC)$ over the basis of integral homology $\mathcal{H}_{\bullet}(\mathcal{C},\ZZ)$ associated with the boundary states. Even when a direct geometric interpretation is not known, the emergence of generalized symmetries should enforce nontrivial algebraic relations among these periods, which should lead to significant simplifications in their structure. 

Section \ref{s:proposal} constitutes the core of the article, where we present the proposal and discuss its properties, consequences, and possible modifications. In Section \ref{s:examples} we explicitly apply the construction and the corresponding definitions to the case of sigma models with target space given by an elliptic curve (Section \ref{s:elliptic_curves}) and a $K3$ surface (Section \ref{s:K3_surfaces}).
We conclude in Section \ref{s:conclusions} with some final remarks and an outlook for future investigations.

\newpage
\section{The proposal}
\label{s:proposal}
    We consider (Euclidean, compact, unitary) 2D superconformal field theories $\calC$ with $\CN=(2,2)$ superconformal symmetries at central charges $c=\tilde c=3d$, with $d\in \NN$, and `spectral flow invariant', in the sense explained below. The main examples we have in mind are supersymmetric sigma models with target space a Calabi-Yau manifold (including complex tori) with complex dimension $d$. Recall that the $\CN=2$ superconformal algebra $(\CN=2)_c$
    \begin{equation}
    \begin{split}
        & \left[ L_m , L_n \right] = (m-n)L_{m+n} + \frac{c}{12} (m^3-m) \delta_{m+n,0}\\
        &\left[ L_m , j_n\right] =-nj_{m+n} \\
        & \left[ L_m , G_r^{\pm}\right] =\left( \frac{m}{2}-r\right) G^{\pm}_{m+r} \\
        & \left[ j_m , j_n \right] = m \delta_{m+n,0} \\
        & \left[ j_n, G_r^{\pm}\right] = \pm G^{\pm}_{r+m} \\
        & \left\lbrace G^+_r , G^+_s \right\rbrace = \left\lbrace G^-_r , G^-_s \right\rbrace =0 \\
        & \left\lbrace G^+_r , G^-_s \right\rbrace = 2 L_{r+s} +(r-s)j_{r+s}+ \left( r^2 - \frac{1}{4} \right) \delta_{r+s,0}.
    \end{split}
    \label{e:N=2SCA}
\end{equation}
admits a family of spectral flow automorphisms parametrised by $\eta\in \RR$:
\begin{equation}
    \begin{split}
        & L_n \quad \mapsto \quad L'_n = L_n + \eta j_n + \frac{\eta^2}{6} c \delta_{n,0} \\
     & j_n \quad \mapsto \quad j'_n = j_n + \frac{c}{3} \eta \delta_{n,0} \\
             & G_r^{\pm} \quad \mapsto \quad G'^{\pm}_r= G^{\pm}_{r + \eta} \\
    \end{split}
\end{equation}
     For $\eta\in \frac{1}{2}+\ZZ$, the automorphism relates the NS and the Ramond representations, while for integral $\eta\in \ZZ$ it relates representations in the same sector (NS or R). We assume that the chiral and antichiral algebras $\CA$ and $\tilde \CA$ of the theory $\calC$ contain the extension of the $\CN=2$ algebra by NS-NS (anti-)holomorphic fields $V_{(\eta, \bar{\eta})}=V_{(\pm 1,0)}(z)$ and $\tilde{V}_{(\eta, \bar{\eta})}=\tilde V_{(0,\pm 1)}(\bar z)$ (the NS-NS spectral flow generators) with conformal weights and $U(1)$ charges $(h,q;\tilde h,\tilde q)=(\frac{c}{6},\pm \frac{c}{3};0,0)$ and $(h,q;\tilde h,\tilde q)=(0,0;\frac{c}{6},\pm \frac{c}{3})$:
     \be \CA\supseteq (\CN=2)_c^{s.f.}\supset (\CN=2)_c\ee
     where 
     $$ (\CN=2)_c^{s.f.}:\quad \text{algebra generated by }T(z),G^\pm(z),j(z)\text{ and spectral flow operators }V_{(\pm 1,0)}(z)\ .
     $$
     The OPE of any field with $V_{(\eta,\tilde\eta)}$ induces an integral spectral flow of either the holomorphic or anti-holomorphic $\CN=2$ algebra by $(\eta,\tilde \eta)=(\pm 1,0)$ or $(\eta,\tilde \eta)=(0,\pm 1)$. The algebra $(\CN=2)_c^{s.f.}$ is a simple current extension of the $\CN=2$ superconformal algebra, and locality requires the $U(1)$ R-charges (i.e., $j_0,\tilde j_0$-eigenvalues) of all NS-NS operators to be integral.   
     We also assume that the R-R sector is related to the NS-NS one by half-integral spectral flow with $(\eta,\tilde \eta)=(\pm 1/2,\pm 1/2)$. The four R-R operators $V_{(\pm 1/2,\pm 1/2)}$ that are related to the NS-NS vacuum are the generators of this spectral flow.

    Although we will formulate all our conjectures in the context of 2D CFT $\calC$, our motivation and inspiration comes from considering type II superstring compactified on $\calC\times \CM_{1,9-2d}$, for some space-time manifold $\CM_{1,9-2d}$. This means that one should include also the R-NS and NS-R sectors of $\calC$, related to the NS-NS by $(\eta,\tilde \eta)=(\pm 1/2,0)$ and $(0,\pm 1/2)$ spectral flow,  tensor each sector of $\calC$ with the corresponding sector of $\CM_{1,9-2d}$, and implement a suitable GSO projection on the complete theory. The BRST operator defining the physical closed string states is constructed in terms of the holomorphic and antiholomorphic $\CN=1$ supercurrent
    \be\label{Neq1} G(z):=G^+(z)+G^-(z)\qquad \tilde G(z):=\tilde G^+(z)+\tilde G^-(z)\ ,
    \ee of $\calC$.
    The resulting model has extended space-time supersymmetries, with gravitini corresponding to the R-NS and NS-R spectral flow operator, and admits BPS D-branes that are charged with respect to the massless R-R fields. 

    The massless R-R fields in the compactified space-time correspond to R-R ground states (i.e. with $h=\tilde h=\frac{c}{24}=\frac{d}{8}$) in the internal SCFT $\calC$.  The Hilbert space of such states decomposes into eigenspaces for the $U(1)$ R-charges $j_0,\tilde j_0$, with eigenvalues $(q,\tilde q)$ satisfying $q,\tilde q\in \frac{c}{6}+\ZZ=\frac{d}{2}+\ZZ$, $|q|,|\tilde q|\le d/2$. It is useful to denote the space of R-R ground states as
\be \CH^\bullet(\calC,\CC):=\left\{|v\rangle\in \Hh_{R-R}\mid L_0|v\rangle=\tilde L_0|v\rangle=\frac{c}{24}|v\rangle\right\}=\bigoplus_{r,s=0}^d \CH^{r,s}(\calC,\CC)
\label{e:H(C,C)}\ee where the $\CH^{r,s}(\calC,\CC)$ are the $U(1)$ R-charges eigenspaces
\be \CH^{r,s}(\calC,\CC):=\left\{|v\rangle\in \Hh_{R-R}\mid L_0|v\rangle=\tilde L_0|v\rangle=\frac{c}{24}|v\rangle,\ j_0|v\rangle=(-\frac{d}{2}+r)|v\rangle,\  \tilde j_0|v\rangle=(\frac{d}{2}-s)|v\rangle,\right\}\ , \label{e:grading_GS_space}
\ee so that $r,s=0,\ldots, d$. The dimensions of these eigenspaces
\be h^{r,s}:=\dim_\CC \CH^{r,s}(\calC,\CC)
\ee coincide with the Hodge numbers of the corresponding Calabi-Yau target space. Roughly speaking, the spaces $\CH^{r,d-r}(\calC,\CC)$, $r=0,\ldots,d$ correspond to the Hodge components of the middle cohomology of the target space, while  $\CH^{r,-r}(\calC,\CC)$ correspond to the Hodge components of the middle cohomology of the mirror Calabi-Yau. One always has 
\be h^{0,0}=h^{d,d}=h^{d,0}=h^{0,d}=1\ ,
\ee and the corresponding $1$-dimensional spaces are spanned by the R-R spectral flow operators $V_{(\pm \frac{1}{2},\pm \frac{1}{2})}$.\\
The CPT operator $\Theta$ defines an antilinear involution on $\CH^\bullet(\calC,\CC)$ that exchanges the components with opposite charges 
\be \Theta: \CH^{r,s}(\calC,\CC)\to \CH^{d-r,d-s}(\calC,\CC)\ .
\ee We denote by $\CH^\bullet(\calC,\RR)$ the real subspace of $\CH^\bullet(\calC, \mathbb{C})$ of CPT self-conjugate states
\be \CH^\bullet(\calC,\RR):=\{|v\rangle\in \CH^\bullet(\calC,\CC)\mid \Theta|v\rangle=|v\rangle\}
\ee so that
\be \CH^\bullet(\calC,\CC)= \CH^\bullet(\calC,\RR)\otimes\CC\ .
\ee

    The BPS D-branes in the full ten dimensional type II string theory can be obtained from particular kind of boundary conditions $\CB_\alpha$ (or, equivalently, boundary states $||\CB\rrangle$) in the SCFT $\calC$. Recall that the boundary states $||\CB\rrangle$ are (non-normalizable) superpositions of closed string states in the Hilbert space $\Hh$ of $\calC$. One of the properties of the BPS boundary states is that they preserve the $\CN=2$ superconformal algebra, i.e. they obey either type-A gluing conditions
\begin{equation}
    \text{type-A}: \quad  \left( L_n- \overline{L}_{-n}\right) \vert\vert \CB \rrangle_\epsilon =0\quad \quad \left( j_n- \overline{j}_{-n}\right) \vert\vert \CB \rrangle_\epsilon =0  \, \, \, , \quad \quad \left( G_r^{\pm} +i\epsilon \overline{G}^{\mp}_{-r}\right) ||\CB\rrangle_\epsilon =0 
\end{equation}
or type-B gluing conditions
\begin{equation}
    \text{type-B}: \quad \left( L_n- \overline{L}_{-n}\right) \vert\vert \CB \rrangle_\epsilon =0\quad \quad \left( j_n+ \overline{j}_{-n}\right) \vert\vert B\rrangle_\epsilon =0  \, \, \, , \quad \quad \left( G_r^{\pm} +i\epsilon \overline{G}^{\pm}_{-r}\right) \vert\vert B\rrangle_\epsilon =0 ,
\end{equation} where $\epsilon\in\{\pm 1\}$.\\
In the full $10$-dimensional theory, the GSO projection forces one to consider a combination of boundary states with different signs $\epsilon$ for the gluing conditions on the supercurrents. Furthermore, the BPS condition implies that one should take a suitable linear combination of NS-NS and R-R sector boundary states $||\CB\rrangle_{\epsilon,NS-NS}$ and $||\CB\rrangle_{\epsilon,R-R}$, in such a way that the spectrum of open strings suspended between two D-branes of the same kind is supersymmetric, and in particular does not contain any tachyon. This ensures that the BPS brane is indeed stable, as expected. More precisely, the R-R contribution $||\CB\rrangle_{R-R}$ and the NS-NS contribution $||\CB\rrangle_{NS-NS}$ must be related by spectral flow with $(\eta,\tilde\eta)=(\frac{1}{2},\frac{1}{2})$.

The R-R charge of a boundary state $||\CB\rrangle$ with respect to a massless field corresponding to the R-R ground state $|v\rangle\in \CH^\bullet(\calC,\CC)$ is
\be \llangle \CB||v\rangle=\langle v(0)\rangle_{\Delta_1,\CB}\ ,
\ee 
and is given by the $1$-point function on the disk $\Delta_1$ with unit radius for the operator $v(z)$ inserted at the origin $z=0$ and with boundary condition $\CB$ on the circle $S^1=\partial \Delta_1$. One can define the charge of a boundary $\CB$ as an element $\Gamma_\CB$ in the space $\CH^\bullet(\calC,\CC)^*$ dual to the space of R-R ground fields, given by \begin{align*} \Gamma_\CB:\CH^\bullet(\calC,\CC)&\to \CC\\ v&\mapsto \Gamma_\CB(v):=\llangle \CB||v\rangle \ .\end{align*} Equivalently, one can define $w_\CB\in \CH^\bullet(\calC,\CC)$ as the ground state component of the R-R boundary state $||\CB\rrangle_{R-R}$, so that
\be \Gamma_\CB(v)=\langle w_\CB|v\rangle \qquad \forall v\in \CH^\bullet(\calC,\CC)\ .
\ee
The set of all (formal) integral linear combinations of charges $\Gamma_\CB\in \CH^\bullet(\calC,\CC)^*$, where $\CB$ is a BPS boundary state, forms a lattice 
\be \CH_\bullet(\calC,\ZZ):=\ZZ\{\Gamma_\CB\mid \CB\text{ BPS boundary}\}\subset \CH^\bullet(\calC,\CC)^*\ , \label{e:BS_lattice}
\ee and the  dual lattice is the $\ZZ$-linear span of all ground states components $w_\CB\in \CH^\bullet(\calC,\CC)$ in the R-R BPS boundary states
\be \CH^\bullet(\calC,\ZZ):=\ZZ\{w_\CB\mid \CB\text{ BPS boundary}\}\subset \CH^\bullet(\calC,\CC)\ .
\ee
For our construction we will also need the rational vector space given by rational linear combinations of the $w_\CB$
\be \CH^\bullet(\calC,\QQ):=\CH^\bullet(\calC,\ZZ)\otimes \QQ\ .
\ee
It is clear that the A-type (resp., B-type) D-branes can only be charged with respect to R-R fields with $q=\tilde q$ (resp., $q=-\tilde q$). Thus, for A-type D-branes, the relevant R-R field form the middle cohomology
\be \text{A-type:}\qquad \CH^{middle}(\calC,\CC)=\oplus_{s=0}^d \CH^{d-s,s}(\calC,\CC)\ ,
\ee while for B-type branes the relevant fields are the ones in the `vertical cohomology' (i.e. the middle cohomology of the mirror theory)
\be \text{B-type:}\qquad \CH^{vert}(\calC,\CC)=\oplus_{s=0}^d \CH^{s,s}(\calC,\CC)\ .
\ee The middle and vertical rational cohomology of $\calC$, defined in terms of BPS D-brane charges, provide examples of rational Hodge structures.\footnote{Notice that for $T^4$, $T^6$, and CY $4$-folds, there are (in general) other R-R fields that are not contained in these two subspaces. In the case of tori, this is due to the fact that these models contain many different copies of the $\CN=2$ superconformal algebra, and the supersymmetric D-branes that preserve these algebras are charged with respect to such R-R fields. For CY $4$-folds, however, it might happen that there are no supersymmetric D-branes charged with respect to the fields in $\CH^{r,s}(\calC,\CC)$ for $(r,s)\in \{(2,1),(1,2),(2,3),(3,2)\}$.}

\vspace{0.15cm}
\noindent
\textbf{Definition:} [Rational/Real Hodge structure]\\
\textit{A rational (real) Hodge structure of weight $w$ is a tuple $\left( V_{\mathbb{Q}}, V_{\mathbb{C}} \right)$ ($\left( V_{\mathbb{R}}, V_{\mathbb{C}} \right)$) of a $\mathbb{Q}$ ($\mathbb{R}$) vector space $V_{\mathbb{Q}}$ ($V_{\mathbb{R}}$) together with a decomposition of its complexification:
\begin{equation}
    V_{\mathbb{C}} = V_{\mathbb{Q}} \otimes_{\mathbb{Q}} \mathbb{C} = \bigoplus_{p+q=w} V_{\mathbb{C}}^{p,q}
\end{equation}
\begin{equation}
    \left( \, \, \, V_{\mathbb{C}} = V_{\mathbb{R}} \otimes_{\mathbb{R}} \mathbb{C} = \bigoplus_{p+q=w} V_{\mathbb{C}}^{p,q} \, \, \, \right),
\end{equation}
such that
\begin{equation}
    \overline{V^{p,q}_{\mathbb{C}}} =V^{q,p}_{\mathbb{C}}.
\end{equation}
}
The dimension of the graded pieces $h^{p,q}=dim_{\mathbb{C}} V^{p,q}_{\mathbb{C}}$ are called the Hodge numbers, and the array $\left( h^{w,0}, h^{w-1,1},...,h^{0,w}\right)$ identifies the Hodge type of the structure. It is clear that both the middle cohomology and the vertical cohomology are Hodge structures of weight $d$, by setting \be \text{Middle Hodge structure:}\qquad V^{r,d-r}_\QQ=\CH^{r,d-r}(\calC,\QQ)\ee and \be \text{Vertical Hodge structure:}\qquad  V^{r,d-r}_\QQ=\CH^{r,r}(\calC,\QQ)\ ,\ee respectively.

\vspace{0.15cm}
\noindent
\textbf{Definition:}[Hodge substructure]\\
\textit{An Hodge substructure of a rational (real) Hodge structure $\left( V_{\mathbb{Q}}, V_{\mathbb{C}} \right)$ ($\left( V_{\mathbb{R}}, V_{\mathbb{C}} \right)$) of a $\mathbb{Q}$ ($\mathbb{R}$) is a $\mathbb{Q}$ ($\mathbb{R}$) vector subspace $W_{\mathbb{Q}} \subset V_{\mathbb{Q}}$ ($W_{\mathbb{R}} \subset V_{\mathbb{R}}$) compatible with the Hodge decomposition of $V_{\mathbb{Q}}$ ($V_{\mathbb{R}}$), namely
\begin{equation}
    W_{\mathbb{C}} = W_{\mathbb{Q}} \otimes_{\mathbb{Q}} \mathbb{C} = \bigoplus_{p+q=w} W_{\mathbb{C}} \cap V^{p,q}_{\mathbb{C}}
\end{equation}
\begin{equation}
    \left( \, \, \, W_{\mathbb{C}} = W_{\mathbb{Q}} \otimes_{\mathbb{Q}} \mathbb{C} = \bigoplus_{p+q=w} W_{\mathbb{C}} \cap V^{p,q}_{\mathbb{C}} \, \, \, \right).
\end{equation}
}

An Hodge structure that has no non-trivial Hodge substructures is called simple, otherwise is called reducible.

\vspace{0.15cm}
\noindent
\textbf{Definition:}[Polarization]\\
\textit{A rational (real) Hodge structure is said to be polarized if there exists a polarization, namely a non-degenerate bilinear form $Q: V_{\mathbb{Q}} \times V_{\mathbb{Q}}\mapsto \QQ$ ($Q: V_{\mathbb{R}}\times  V_{\mathbb{R}}\mapsto \RR$) that extends by linearity to $V_{\mathbb{C}}$ and satisfies the following two properties:
\begin{equation}
\begin{split}
& (i) \quad \textsc{orthogonality} \quad Q \left( V^{p,q} , V^{r,s} \right)=0 \quad \text{for} \quad p \neq s, q \neq r ; \\
& (ii) \quad \textsc{non-degeneracy} \quad i^{p-q} Q \left( v, \overline{v} \right) >0 \quad \text{for} \quad v \in V^{p,q}, \quad v \neq 0 .   \\
\end{split}
\label{e:polarization_properties}
\end{equation} 
}\\
Notice that if an Hodge structure is polarized, then all its substructures are polarized as well. Furthermore, a polarized Hodge structure decomposes into an orthogonal sum of irreducible polarized Hodge substructures. 
A polarization $Q$ on $\CH^\bullet(\calC,\QQ)$ can be defined by considering the open string Witten index between two supersymmetric boundary conditions $\CB,\CB'$ with charges $w_\CB,w_{\CB'}\in \CH^\bullet(\calC,\ZZ)$
\be Q(w_\CB,w_\CB'):=\langle w_\CB|(-1)^{\tilde F}|w'_\CB\rangle={}_{RR}\llangle \CB||(-1)^{\tilde F} q^{L_0+\bar L_0-\frac{c+\tilde c}{24}}||\CB'\rrangle_{RR} \label{e:Witten_index}
\ee and then extending by $\QQ$-linearity to $\CH^\bullet(\calC,\QQ)$.

Let us now consider the generalized symmetries of the model $\calC$, i.e. the topological defect lines. We recall some basic facts about topological lines, and refer to the literature for more details \cite{Frohlich:2004ef, Frohlich:2009gb, Frohlich:2006ch, Chang:2019ccl, Bhardwaj:2017xup, Thorngren:2019rtw, Thorngren:2021rtw}. Consider the CFT on a cylinder $S^1\times \RR$, where $S^1$ is the space direction and $\RR$ the Euclidean time direction. Then, with each topological line $\CL$ is associated a linear operator $\hat\CL:\CH\to \CH$ and a $\CL$-twisted sector $\CH_\CL$, that correspond to inserting the defect, respectively, along the $S^1$ at fixed time or along the time line $\RR$. The fusion of two defects $\CL_1\CL_2$ is obtained by moving two lines close to each other so as to coincide. A second operation is given by superposition of defects $\CL_1+\CL_2$. We have \be \widehat{\CL_1\CL_2}=\hat\CL_1\hat\CL_2,\qquad  \widehat{\CL_1+\CL_2}=\hat\CL_1+\hat\CL_2\ ,\ee as well as \be \CH_{\CL_1\CL_2}=\CH_{\CL_1}\otimes\CH_{\CL_2}\ ,\qquad  \CH_{\CL_1+\CL_2}=\CH_{\CL_1}\oplus\CH_{\CL_2}\  ,\ee (for a suitable notion of fusion product of spaces). A defect $\CL$ is called simple if it cannot be written as the superposition of other defects. The identity defect $\CI$ is such that $\hat\CI$ is the identity operator and $\CH_\CI=\CH$. For each defect line $\CL$, there is a dual defect $\CL^*$ obtained by changing the orientation of the line.  One has $\CL\CL^*=n\CI+\ldots$ with $n\in\ZZ_{>0}$, and $n=1$ if and only if $\CL$ is simple. A defect is called invertible if $\CL\CL^*=\CI$; in this case, the operator $\hat\CL$ is unitary, and $\hat\CL^*=\hat\CL^{-1}$. The eigenvalue of $\hat\CL$ on the vacuum state is called the quantum dimension; we denote it by $\langle \CL\rangle$. In a unitary CFT with a unique vacuum, it is always a real number with $\langle \CL\rangle\ge 1$, and the equality holds if and only if $\CL$ is a simple invertible defect. Furthermore, one has
\be \langle \CL^*\rangle=\langle \CL\rangle\ ,\qquad \langle \CL_1\CL_2\rangle=\langle \CL_1\rangle\langle\CL_2\rangle\qquad \langle \CL_1+\CL_2\rangle=\langle \CL_1\rangle+\langle\CL_2\rangle\ .
\ee From a mathematical perspective, topological defects correspond to objects in a fusion, or more general, tensor category (see \cite{Bhardwaj:2018blt, Chang:2019ccl, Thorngren:2019rtw, Thorngren:2021rtw} for an introduction to fusion categories in two dimensions and \cite{Etingof:2015} for tensor categories).

When a topological defect $\CL$ is moved across a local operator $\phi(z)$, the latter can get transformed into a non-local operator $\tilde\phi(z)$ attached to a line $\CL\CL^*$. We say that a certain operator $\phi(z)$ is preserved by $\CL$ (or that $\phi(z)$ and $\CL$ are transparent to each other) if moving $\CL$ across $\phi(z)$ does not modify $\phi(z)$. More generally, if we have a subCFT generated by a subset of local operators that is closed under OPE (on the sphere), we say that a simple defect $\CL$ acts invertibly on such a subCFT if moving $\CL$ across the insertion of a $\phi(z)$, the latter gets transformed into a \emph{local} operator in the same subCFT, in a way compatible with the OPE (i.e., $\CL$ acts by a standard invertible automorphism of the subCFT).

After this general introduction, let us now discuss the topological defect lines on the superconformal field theory $\calC$. The category $\TDL$ we are mostly interested in contains all topological defect $\CL$ in $\calC$ that satisfy the following properties:
\begin{enumerate}
    \item every $\CL\in \TDL$ preserves (is transparent to) the holomorphic and anti-holomorphic $\CN=2$ superconformal algebra.
    \item The simple defects $\CL$ act invertibly on the subCFT generated by the R-R spectral flow operators $V_{(\pm 1/2,\pm 1/2)}$, as well as on the holomorphic and anti-holomorphic algebras $(\CN=2)_c^{sf}$ extended by the NS-NS spectral flow operators $V_{(\pm 1,0)}$ and $V_{(0,\pm 1)}$.
\end{enumerate}

These conditions imply that moving a simple defect $\CL$ across a (anti-)holomorphic operator $\phi(z)$ in the (anti-)chiral algebra $(\CN=2)_c^{sf}$, such operator gets transformed by an automorphism of $(\CN=2)_c^{sf}$ that fixes the subalgebra $\CN=2$. The only way such a automorphism can act on the NS-NS spectral flow operators $V_{(\pm 1,0)}$ and $V_{(0,\pm 1)}$ is via multiplication by a phase
\be\label{ActOnSpecFlow1} V_{(\pm 1,0)}\mapsto g_L(\CL)^{\pm 1}V_{(\pm 1,0)}\ ,\qquad V_{(0,\pm 1)}\mapsto g_R(\CL)^{\pm 1}V_{(0,\pm 1)}\ ,
\ee for some phases $(g_L(\CL),g_R(\CL))\in U(1)\times U(1)$. Compatibility with the OPE implies  that when one moves $\CL$ across a R-R spectral flow operators $V_{(\pm 1/2,\pm 1/2)}(z)$, the latter gets transformed as
\be\label{ActOnSpecFlow2} V_{(\eta,\tilde\eta)}\mapsto g_L(\CL)^{\eta}g_R(\CL)^{\tilde\eta}V_{(\eta,\tilde\eta)}\ ,\qquad \eta,\tilde\eta\in \{\pm 1/2\}\ .
\ee At the level of the corresponding states, this means that the operator $\hat\CL$ associated with a simple defect $\CL\in \TDL$ acts by
\be \hat\CL|\phi\rangle=\langle \CL\rangle |\phi\rangle\ ,
\ee on any state $|\phi\rangle$ corresponding to an operator in the preserved $\CN=(2,2)$ superconformal algebra, and by
\be \hat\CL|V_{(\eta,\tilde\eta)}\rangle =g_L(\CL)^{\eta}g_R(\CL)^{\tilde\eta}\langle \CL\rangle|V_{(\eta,\tilde\eta)}\rangle\ ,
\label{e:L_action_sf_states}\ee on any NS-NS or R-R spectral flow state $|V_{(\eta,\tilde\eta)}\rangle$. 

 The motivation for considering a tensor category of defects with these properties comes again from string theory. As anticipated above, we look for topological defects in the full CFT $\calC\times \CM_{1,9-2d}$ that act only on the internal factor $\calC$, that preserve the physical state condition for fundamental strings, and relate BPS D-branes to BPS D-branes within the same string model. This forces us to consider defects in $\calC$ that are transparent to the $\CN=1$ supercurrent $G(z)$ in eq.\eqref{Neq1} that contributes to the string BRST operator,  and that acts invertibly on the NS-R and R-NS spectral flow operators implementing space-time supersymmetry. The chiral algebra generated by $G(z)$ and spectral flow operators is $(\CN=2)^{s.f.}$, and the group of automorphisms fixing $G(z)$ is (generically) $O(2)$. It turns out that the holomorphic and antiholomorphic $SO(2)\subset O(2)$ subgroups of automorphisms fix the whole $\CN=(2,2)$ superconformal algebra, while they act as in \eqref{ActOnSpecFlow1} and \eqref{ActOnSpecFlow2} on the spectral flow operators. The automorphisms of the $(\CN=2)^{sf}$ algebra in $O(2)\setminus SO(2)$ include mirror symmetry, that for odd dimension $d$ map type IIA to type IIB GSO projection; therefore, they do not always relate fundamentals strings and D-branes within the same string model. In this article, we focus on the category $\TDL$ of defects acting on both the holomorphic and antiholomorphic $(\CN=2)^{s.f}$ algebras by automorphisms in the subgroups $SO(2)\subset O(2)$. While it might make sense in some particular CFTs $\calC$ to consider slightly more general defects,  this choice allows us to formulate general conjectures in a uniform way.  

    In general, the fusion of a topological defect $\CL$  with a boundary $\CB$ provides a new boundary $\CL\otimes\CB$, corresponding to the boundary state $\hat\CL||\CB\rrangle$. If $\CB$ is a BPS D-brane with charge $w_\CB\in \CH^\bullet(\calC,\ZZ)$ and $\CL$ is a topological defect in $\TDL$,  then $\CL\otimes\CB$ is again a BPS D-brane with charge $\hat\CL(w_\CB)\in \CH^\bullet(\calC,\ZZ)$. This means that each $\CL\in \TDL$ induces a $\ZZ$-linear map 
    \be
    \EndL:=\hat\CL_{\rvert \CH^\bullet(\calC,\ZZ)}: \CH^\bullet(\calC,\ZZ)\to \CH^\bullet(\calC,\ZZ)\ , \label{e:lattice_end}\ee i.e. an endomorphism $\EndL \in \End(\CH^\bullet(\calC,\ZZ))$. Furthermore, the fact that $\CL$ preserves the holomorphic and anti-holomorphic $U(1)$ currents implies that the $\CC$-linear extension $\EndL_\CC$, as a $\CC$-linear map on $\CH^\bullet(\calC,\CC)$, preserves each subspace $\CH^{r,s}(\calC,\CC)$
    \be \EndL_\CC (\CH^{r,s}(\calC,\CC))\subseteq \CH^{r,s}(\calC,\CC)\ .
    \ee
In mathematics, lattice endomorphisms of the form \eqref{e:lattice_end} that preserve the grading of the underlying complex cohomology are referred to as \emph{Hodge endomorphisms}.

\vspace{0.15cm}
\noindent
\textbf{Definition:}[Hodge endomorphism]\\
\textit{Let $(V_{\mathbb{Q}}, V_{\mathbb{C}})$ be a rational Hodge structure of weight $w$. A Hodge endomorphism is a vector space endomorphism
\begin{equation}
    \varphi \, \, \, : \, \, V_{\mathbb{Q}} \, \, \mapsto \, \, V_{\mathbb{Q}}
\end{equation}
which complex extension preserves the Hodge decomposition:
\begin{equation}
    \varphi_{\mathbb{C}} \left( V_{\mathbb{C}}^{p,q} \right) \,  \subseteq \, V^{p,q}_{\mathbb{C}}.
\end{equation}
}
\noindent
Thus, the restrictions $\EndL:=\hat\CL_{\rvert \CH^\bullet(\calC,\ZZ)}$ of the linear operators $\hat\CL$ to the lattice of ground states components of the R-R BPS boundary states are examples of these Hodge endomorphisms. We use these endomorphisms to define Hodge loci in the CFTs moduli space\footnote{Here we are using an extension of the notion of Hodge loci from the geometric moduli space of Calabi–Yau manifolds to the moduli space of $\mathcal{N}=(2,2)$ SCFTs with target Calabi-Yau. In the geometric setting, we call Hodge loci the subspaces $HL (\mathcal{M}_{CY_d}, H^d)$ and $HL (\mathcal{M}_{CY_d}, H^{\otimes})$ of $\mathcal{M}_{CY_d}$ corresponding respectively to non-trivial integral or rational Hodge classes in $H^d(X, \mathbb{C})$ and to non-trivial integral or rational Hodge tensors. Non-trivial integral/rational classes/tensors means integral/rational classes/tensors that do not exists already at generic points of the moduli space.}.

\begin{definition}
\label{d:Hodge_loci}
    We define the \textbf{Hodge loci} $HL \left( \mathcal{M}_{\mathcal{N}=2,2}, \TDL \right)$ as the subspaces of the moduli space $\mathcal{M}_{\mathcal{N}=2,2}$ of spectral-flow invariant $\mathcal{N}=(2,2)$ SCFTs, where non-trivial Hodge endomorphisms induced by topological defects lines $\mathcal{L} \in \TDL$, satisfying properties $(1)$ and $(2)$, arise.
\end{definition}

It is therefore natural to consider the space of all such endomorphisms at a given point in the moduli space.

\vspace{0.15cm}
\noindent
\textbf{Definition:}[Algebra of endomorphisms]\\
\textit{The $\QQ$-vector space of Hodge endomorphisms equipped with composition forms a $\mathbb{Q}$-algebra:
\begin{equation}
    \End_{Hdg} (V_{\mathbb{Q}}) = \left\lbrace \varphi \in \mathrm{Hom} (V_{\mathbb{Q}}, V_{\mathbb{Q}}) \vert \varphi_{\mathbb{C}} (V^{p,q}_{\mathbb{C}}) \subseteq V^{p,q}_{\mathbb{C}} \right\rbrace.
\end{equation}
}

\noindent
Similarly, the $\QQ$-vector space generated by the endomorphisms $\EndL$, for all $\CL$ in $\TDL$ (or more generally, in a tensor subcategory $\TDL'$ of $\TDL$) is an algebra
\be \End_{\TDL'}:=\QQ\{\EndL\in \End( H^\bullet(\calC,\QQ))\mid  \EndL_\CC=\hat\CL_{\rvert \CH^\bullet(\calC,\CC)}\text{ for some }\CL\in \TDL' \}\ .
\ee Duality of defect $\CL\mapsto \CL^*$ induces an involution in this algebra 
\be \EndL\mapsto \bar\EndL\ ,
\ee where $\bar\EndL_\CC=\hat\CL^*_{\rvert \CH^\bullet(\calC,\CC)}$. The involution is the adjoint with respect to the polarization $Q$, i.e. it satisfies
\be Q(v,\EndL w)=Q(\bar\EndL v ,w)\ .
\ee This follows by considering the CFT amplitude for a bounded cylinder $S^1\times [a,b]$ with conditions $\CB$, $\CB'$ at the boundaries, the defect $\CL$ wrapping the circle $S^1$, and no insertion of local operators. By moving the defect $\CL$ from one boundary to the other, one gets \be
{}_{RR}\llangle \CB||(-1)^{\tilde F} q^{L_0+\bar L_0-\frac{c+\tilde c}{24}}||\CL\otimes\CB'\rrangle_{RR}={}_{RR}\llangle \CL^*\otimes \CB||(-1)^{\tilde F} q^{L_0+\bar L_0-\frac{c+\tilde c}{24}}||\CB'\rrangle_{RR}\ ,   \ee from which the corresponding relation for the polarization follows.

The algebra of Hodge endomorphisms may, in certain cases, admit an embedding of a number field that introduces an additional module structure on the space of R-R boundary states. This is formalized in the following definition.

\vspace{0.15cm}
\noindent
\textbf{Definition:}[E-multiplication]\\
\textit{We say that an irreducible Hodge structure $(V_{\mathbb{Q}}, V_{\mathbb{C}})$ has an $E$-multiplication if there exists an embedding of the number field $E$\footnote{Finite dimensional extension of $\mathbb{Q}$ of degree $dim \left[E: \mathbb{Q} \right]$.} in $\End_{Hdg} (V_{\mathbb{Q}})$:
\begin{equation}
    \xi \, \, \, : \, \, E \, \, \hookrightarrow \, \, \End_{Hdg} (V_{\mathbb{Q}})
    \label{e:E-multiplication}
\end{equation}
which is a $\mathbb{Q}$-morphism.
}

In other words if a rational Hodge structure admit $E$-multiplication, we can associate to each number $e \in E$ an Hodge endomorphism $\xi (e)$ through the map \eqref{e:E-multiplication}: this induces an action of the number field $E$ on $V_{\mathbb{Q}}$, and $V_{\mathbb{C}}$ by complex extension, which makes $V_{\mathbb{Q}}$ an $E$-vector space. In particular, called $n= \dim_E V_{\mathbb{Q}}$ the dimension of $V_{\mathbb{Q}}$ as an $E$-vector space, we have that its dimension as a $\mathbb{Q}$-vector space is
\begin{equation}
    \dim_{\mathbb{Q}} V_{\mathbb{Q}}= \, n \cdot \left[ E: \mathbb{Q}\right].
    \label{e:dimVQ_Qvectspace}
\end{equation}

Let $K:=\xi(E)\subseteq End_{Hdg} (V_{\mathbb{Q}})$ be a subalgebra of the Hodge endomorphism algebra of $(V_{\mathbb{Q}}, V_{\mathbb{C}})$ admitting an embedding of the number field $E$. This means that we can think about $K$ as a number field over $\mathbb{Q}$. To specify which kind of extension it defines we consider its real embeddings
\begin{equation}
    \rho_1 , \, \rho_2 , \,... \, , \rho_r \, \, : K \, \hookrightarrow \, \mathbb{R}
\end{equation}
and its complex ones
\begin{equation}
    \sigma_1 , \, \overline{\sigma}_1 , \,... \, , \sigma_s , \, \overline{\sigma}_s \, \, : K \, \hookrightarrow \, \mathbb{C}.
\end{equation}
The dimension of the field $K$ becomes
\begin{equation}
    \left[ K : \mathbb{Q}\right] \, = \, r +2s.
\end{equation}
Let us introduce the following two types of number fields depending on the kinds of embeddings.

\vspace{0.15cm}
\noindent
\textbf{Definition:}[Real Field]\\
\textit{A number field $K$ is called totally real if $s=0$.}

\vspace{0.15cm}
\noindent
\textbf{Definition:}[Complex Multiplication Field and Algebra]\\
\textit{A number field $K$ is a Complex Multiplication (CM) field if it contains a subfield $K_0 \subset K$ that is totally real and $K / K_0$ is a purely imaginary quadratic extension. A Complex Multiplication algebra is a finite product $K_1\times \ldots\times K_n$ of CM fields $K_i$.}

To have a CM field means that for any real embedding $\rho : K_0 \hookrightarrow \mathbb{R}$, the corresponding extension to $K$ generates two complex conjugate embedding into $\mathbb{C}$. Then, if $K$ is a CM field, it can be obtained by adding an imaginary square root to a totally real subfield $K_0$:
\begin{equation}
     \exists \, \alpha \, \in K_0 \, \vert \rho_i (\alpha) \in \mathbb{R}_{<0} \, \forall \rho : K_0 \hookrightarrow \mathbb{R} \, \,  \text{and} \,\, K= K_0 (\sqrt{ \alpha}).
\end{equation}

\vspace{0.15cm}

We say that the irreducible Hodge structure $(V_{\mathbb{Q}}, V_{\mathbb{C}})$ has complex (real) multiplication if it has $K$-multiplication and $K$ is a CM (totally real) field. Notice that, given \eqref{e:dimVQ_Qvectspace}, if $\dim_{\mathbb{Q}} V_{\mathbb{Q}} =1 \, \text{mod} \, 2$, then $K$ must be totally real. Indeed, if $K$ were a CM field, then $\left[ K: \mathbb{Q}\right]$ would be even, and $\dim_{\mathbb{Q}} V_{\mathbb{Q}}$ would be even as well, contradicting the assumption. On the other hand, if $n=\dim_K V_{\mathbb{Q}}=1$, then $\dim_{\mathbb{Q}} V_{\mathbb{Q}}= \left[ K : \mathbb{Q} \right]$ and $K$ is a CM field. This means that in this situation the action of $K$ is sufficient to generate the whole $V_{\mathbb{Q}}$ from a single nonzero vector. 

If the Hodge structure $\left( V_{\mathbb{Q}}, V_{\mathbb{C}}\right)$ is not irreducible, but it can be decomposed in finitely many irreducible substructures $W^{\alpha}_{\mathbb{Q}}$, such that:
\begin{equation}
    V_{\mathbb{Q}} = \bigoplus_{\alpha \in \mathcal{I}} W^{\alpha}_{\mathbb{Q}} \quad \quad \vert \quad \quad W_{\mathbb{C}} = \bigoplus_{p+q=w} W_{\mathbb{C}}^{\alpha, (p,q)},
    \label{e:decomposition_irreducible_subs}
\end{equation}
we say that $V_{\mathbb{Q}}$ is of complex multiplication type if each irreducible component is of CM type. This means that for each $\alpha$, the Hodge structure $W^{\alpha}_{\mathbb{Q}}$ admits multiplication by a CM field $K^{\alpha}$.

In the CFT $\calC$, the Hodge structure on $\CH^\bullet(\calC,\CC)$ in general decomposes as
\be \CH^\bullet(\calC,\QQ)=\oplus_{\alpha\in \mathcal{I}} W^\alpha_\QQ\ ,
\ee where $\{(W^\alpha_\QQ,W^\alpha_\CC)\}_{\alpha\in \mathcal{I}}$ is a set of irreducible Hodge substructures, that are mutually orthogonal with respect to the polarization induced by the open string Witten index. In CFT language, the category $\TDL$ of topological defects lines satisfying properties $(1)$ and $(2)$ realizes an algebra of endomorphisms on $\CH^{\bullet}(\mathcal{C}, \mathbb{Z})$. For each irreducible component $W^\alpha_\CC$, we can consider a tensor subcategory $\TDL_\alpha$ of $\TDL$, which defects preserve $W^\alpha_\CC$, in the sense that
\be \hat\CL(W^\alpha_\CC)\subset W^\alpha_\CC\ ,\qquad \forall \CL\in \TDL_\alpha\ .
\ee With each such subcategory $\TDL_\alpha$ is associated a $\QQ$-algebra 
\be \End_{\TDL_\alpha}=\QQ\{\EndL\in \End_{Hdg}(W^\alpha_\QQ)\mid \EndL_\CC=\hat\CL_{|W^\alpha_\CC}\text{ for some }\CL\in \TDL_\alpha\}
\ee
of Hodge endomorphisms of the irreducible substructure $W^\alpha$. We therefore propose the following definition of complex multiplication type in the CFT setting.
\begin{definition}
\label{d:CM_cft}
    We say that $\calC$ has \textbf{complex multiplication} iff for each irreducible Hodge substructure $W^\alpha_\QQ$ in $H^\bullet(\calC,\QQ)=\oplus_{\alpha\in\mathcal{I}}W^\alpha_\QQ$, there exists a tensor subcategory $\TDL_\alpha$ of $\TDL$, 
    such that the corresponding algebra of Hodge endomorphisms $K_\alpha:=\End_{\TDL_\alpha}$ is a CM field  with the largest possible $\QQ$-dimension $\dim_\QQ \End_{\TDL_\alpha}=\dim_\QQ W^\alpha_\QQ$.
\end{definition}


Let us discuss the properties of this definition and some possible modifications.
\begin{enumerate}
    \item While in the standard algebraic geometry one typically considers rational (or integral) Hodge structures on cohomology groups of fixed weight $w$ and polarized by the cup product, in generalized geometry, necessary in string theory compactifications, one must instead consider polarized 
    mixed-Hodge structures, in which cohomology groups of different weights are treated simultaneously. Roughly speaking, this generalization allows to treat the middle and the vertical cohomologies on an equal footing. Geometrically, this corresponds to treat simultaneously the middle cohomology of a Calabi–Yau manifold and the middle cohomology of its mirror dual. This generalized setting requires additional care in handling the decomposition of the full rational (or integral) structure into irreducible Hodge substructures, as is necessary, for example, in the identification of Complex Multiplication points. We will discuss this in detail in the case of elliptic curves (Section \ref{s:elliptic_curves}) and K3 surfaces (Section \ref{s:K3_surfaces}). Note that considering the full mixed Hodge structure, rather than only the individual fixed-weight cohomologies as in standard geometry, is not merely an abstract mathematical generalization, but is in fact required by string theory. Indeed, the presence of the B-field leads to a moduli space that is more naturally described in terms of generalized Calabi–Yau structures \cite{Hitchin:2003hin} (see also \cite{huybrechts:2004dhu} for K3 surfaces), rather than in terms of separate symplectic and complex structures. 
    \item In the case of non-linear sigma models on elliptic curves (see  section \ref{s:elliptic_curves}) and CY $3$-folds with full $SU(3)$ holonomy, the space $\Hh^\bullet(\calC,\CC)$ is the orthogonal sum of the middle and the vertical cohomology, that form two (not necessarily irreducible) Hodge substructures. In these cases, our definition of CM SCFT applies straightforwardly, while for K3 surfaces, complex tori of dimension at least $4$, and for CY $3$-folds with restricted holonomy, such as $K3\times T^2$, one needs some refinement. The main difference is that, in these cases,  the generic chiral algebra of $\calC$ is larger than just a single $\CN=(2,2)$ superVirasoro. In fact, different BPS boundary states and topological defects might preserve different $\CN=(2,2)$ subalgebras of the full chiral algebra, so that the very definition of the Hodge structure on the space of R-R ground fields becomes ambiguous. A second complication arises for target spaces with even complex dimension, at the special loci in the moduli space where the intersection $\CH^{d/2,d/2}(\calC,\CC)\cap \CH(\calC,\QQ)$ is non-empty. Each non-zero element in this intersection (i.e., a \emph{rational Hodge class}) generates a $1$-dimensional subspace of $\CH(\calC,\QQ)$ that is by itself a trivial irreducible Hodge substructure. Due to these $1$-dimensional irreducible components in the Hodge structure, the corresponding $\calC$ cannot be a CM sigma model, according to our definition above. Nevertheless, the algebra of Hodge endomorphisms in these models can contain large CM fields, so it seems more useful to slightly modify our definition so as to accommodate these cases.  As a prototypical example where both such issues occur, in section \ref{s:K3_surfaces} we propose a suitable definition of complex multiplication for K3 sigma models. We leave the detailed treatment of the other cases (complex tori, $K3\times T^2$, ...) to future work.
    \item Our construction is based on the assumption that the charges of BPS D-branes span the whole space of R-R ground fields. Let us spell this assumption more explicitly. We consider the set of all BPS D-branes in our SCFT. Different elements in this set might preserve different subsets of spacetime supercharges, so that the superposition of such boundary states might not, in general, correspond to a BPS configuration by itself. Then, we consider the set of R-R charge vectors $w_\CB$ associated with these D-branes. Our assumption is that, by taking $\ZZ$-, $\QQ$-, or $\CC$-linear combinations of such charge vectors, the corresponding lattice or vector space has the same dimension as the whole space of R-R ground states. This is sufficient for us, because on one hand we know that fusion with topological defect lines in $\TDL$ yields a permutation on the set of BPS D-branes, and therefore a permutation on the corresponding set of R-R charges. On the other hand, we know that topological defects act linearly on the space of R-R ground states, so the permutation of the BPS brane charges is not a generic one -- it is induced by a linear transformation of the corresponding lattice or vector space.\\
    We believe that this assumption holds for all $\CN=(2,2)$ SCFTs of small central charge $c\le 9$, or at least we do not know of any counterexample. For higher central charges, this assumption might in general fail. For example, consider a non-linear sigma model with target space a smooth Calabi-Yau $4$-fold $X$ with $h^{2,1}\neq 0$. A BPS D-brane charged with respect to a R-R field in a $(2,1)$-Hodge component would have to wrap a supersymmetric $3$-cycle in $X$, but it is known that there are no such cycles \cite{Ooguri:1996oyz}. In SCFT language, this statement follows simply by noticing that such a field would neither satisfy the $A$-type  gluing conditions, nor the $B$-type. So, unless that are different copies of the $\CN=(2,2)$ superconformal algebra, we conclude that all BPS boundary states are neutral with respect to such a field.\footnote{In every $\CN=(2,2)$ sigma model with target space a torus of complex dimension $\ge 2$, there are always some R-R fields that do not satisfy either the $A$-type or $B$-type boundary conditions. In this case, however, one always has multiple inequivalent copies of the $\CN=(2,2)$ superconformal algebra, so by varying which subalgebra is preserved by the gluing conditions, one can always span the whole space of R-R ground fields. The presence of multiple $\CN=(2,2)$ algebras leads to ambiguities in the definition of the Hodge grading; in the present article, we discuss this issue only in the case of K3 sigma models. } If this assumption does not hold, one needs to modify our proposal by restricting to the subspace of R-R ground fields that is spanned by the set of BPS charges.
\end{enumerate}

\section{Examples}
\label{s:examples}

\subsection{Elliptic curves} 
\label{s:elliptic_curves}

In this section, we apply the definition of `CFT complex multiplication' to sigma models with target space an elliptic curve. Recall that the supersymmetric sigma model on $T^2$ is determined by a pair of complex moduli $(\tau,\rho)\in \mathbb{H}\times\mathbb{H}$, where $\tau$ is the complex structure modulus and $\rho$ is the complexified Kahler structure modulus. In this case, mirror symmetry is just T-duality along one of the two circles, where one exchanges type IIB on $(\tau,\rho)$ with type IIA on $(\rho,\tau)$.

The algebra $\CA$ of holomorphic fields is generated by two chiral scalars $X^1,X^2$ and two chiral free fermions $\psi^2,\psi^2$, that can be arranged into complex fields as
\begin{align}
    Z&=\frac{1}{\sqrt{2}}(X^1+iX^2)\ ,& Z^*&=\frac{1}{\sqrt{2}}(X^1-iX^2)\\
    \psi^+&=\frac{1}{\sqrt{2}}(\psi^1+i\psi^2)\ ,& \psi^-&=\frac{1}{\sqrt{2}}(\psi^1-i\psi^2)\ ,
\end{align} with mode expansions
\begin{align}
    \partial Z(z)&=-i\sum_{n\in \ZZ} \alpha_n z^{-n-1}\ ,  & \partial Z^*(z)&=-i\sum_{n\in \ZZ} \alpha^*_n z^{-n-1}\ ,\\
    \psi^+(z)&=\sum_{r} \psi^+_r z^{-r-\frac{1}{2}}\ ,  & \psi^-(z)&=\sum_{r} \psi^-_r z^{-r-\frac{1}{2}}\ ,
\end{align} where $r\in \frac{1}{2}+\ZZ$ for the NS sector and $r\in \ZZ$ for the Ramond sector. The Virasoro, supercurrents and $U(1)$ current of the $(\CN=2)_{c=3}\subset \CA$ superconformal algebra are given by
\begin{align}
    T&=-:\partial Z\partial Z^*:-\frac{1}{2}(:\psi^+\partial\psi^-:+:\psi^-\partial\psi^+:)\\
    G^+&=i\sqrt{2}\psi^+\partial Z^*\\
    G^-&=i\sqrt{2}\psi^-\partial Z\\
    J&=-:\psi^-\psi^+:
\end{align}
The holomorphic NS-NS spectral flow generators $V_{(\pm 1,0)}(z)$, in this model, coincide with the complex free fermions
\be V_{(\pm 1,0)}(z)=\psi^\pm (z)\ .
\ee Therefore, the extension $(\CN=2)^{s.f}_{c=\frac{3}{2}}$ of $(\CN=2)_{c=3}$ by these spectral flow operators is the whole algebra $(\CN=2)^{s.f}_{c=\frac{3}{2}}\equiv \CA$ generated by the free fermions $\psi^\pm$ and free bosons $i\partial Z,i\partial Z^*$.


The primary states with respect to the $\CA\times\tilde\CA$ algebras are $|p,\tilde p\rangle$, where $p,\tilde p\in \CC$ are the complexified left- and right-moving momenta, i.e. the eigenvalues of $\alpha_0$ and $\tilde\alpha_0$
\be \alpha_0|p,\tilde p\rangle =p|p,\tilde p\rangle\qquad \alpha^*_0|p,\tilde p\rangle =p^*|p,\tilde p\rangle
\ee
\be \tilde\alpha_0|p,\tilde p\rangle =\tilde p|p,\tilde p\rangle\qquad \tilde\alpha^*_0|p,\tilde p\rangle =\tilde p^*|p,\tilde p\rangle
\ee (here and in the following $p^*$ denotes the complex conjugate of $p\in \CC$, $\tilde p^*$ is the conjugate of $\tilde p$, while $p\in \CC$ and $\tilde p\in \CC$ are independent complex variable). The state $|p,\tilde p\rangle$ has conformal weights ($L_0$, $\tilde L_0$ eigenvalues) $(h,\tilde h)=(\frac{|p|^2}{2},\frac{|\tilde p|^2}{2})$.

The allowed values of $(p,\tilde p)\in \CC^2$ form an even unimodular lattice $\Gamma^{2,2}_{w-m}\equiv \Gamma^{2,2}$ (the Narain lattice of winding-momentum) with signature $(2,2)$, and bilinear form given by $\|(p,\tilde p)\|^2=|p|^2-|\tilde p|^2$. The precise lattice depends on the moduli $\tau,\rho$, and is given by
\be\label{windmom} \begin{pmatrix}
    p\\ \tilde p
\end{pmatrix}=\frac{i}{\sqrt{2\rho_2\tau_2}}\left[n_2 \begin{pmatrix}
    1\\ 1
\end{pmatrix}+ w_1 \begin{pmatrix}
    \bar \rho\\ \rho
\end{pmatrix} -n_1 \begin{pmatrix}
     \tau\\ \tau
\end{pmatrix}+w_2 \begin{pmatrix}
    \bar\rho \tau\\ \rho\tau
\end{pmatrix}\right]\qquad n_1,w_1,n_2,w_2\in \ZZ
\ee where $n_1,n_2$ are quanta of momenta and $w_1,w_2$ are winding numbers. Equivalently,
\be p=\frac{i}{\sqrt{2\rho_2\tau_2}}\begin{pmatrix}
 \bar\rho & 1
\end{pmatrix}\begin{pmatrix}
    w_2 & w_1\\ -n_1 & n_2
\end{pmatrix}\begin{pmatrix}
    \tau \\ 1
\end{pmatrix}
\ee
\be \tilde p=\frac{i}{\sqrt{2\rho_2\tau_2}}\begin{pmatrix}
 \rho & 1
\end{pmatrix}\begin{pmatrix}
    w_2 & w_1\\ -n_1 & n_2
\end{pmatrix}\begin{pmatrix}
    \tau \\ 1
\end{pmatrix}
\ee If $(p,\tilde p)$ are the vectors associated with $n_1,n_2,w_1,w_2\in \ZZ$, then a direct calculation shows that the metric of signature $(2,2)$ on the lattice is given by
\be\label{latticemetric} |p|^2-|\tilde p|^2=2n_1w_1+2n_2w_2\ .
\ee We will also use the notation
\be |n_k,w_k\rangle \equiv |p,\tilde p\rangle\ ,
\ee for the primary states with winding-momentum $n_1,n_2,w_1,w_2\in \ZZ$.

The model has $4$ R-R ground fields, than can be labeled by their R-symmetry charges with respect to $J_0$ and $\tilde J_0$ as $|\pm,\pm\rangle\equiv|\pm \frac{1}{2},\pm \frac{1}{2}\rangle$. The operators corresponding to such states are exactly the R-R spectral flow generators $V_{(\pm 1/2,\pm 1/2)}$, i.e.
\be \left|\pm \tfrac{1}{2},\pm \tfrac{1}{2}\right\rangle\quad \leftrightarrow \quad V_{(\pm 1/2,\pm 1/2)}(z,\bar z)\ ,
\ee under the state operator correspondence. Thus, the space of RR ground states $\CH^\bullet(\calC,\CC)$ decomposes as
\be \CH^\bullet(\calC,\CC)=\CH^{0,0}(\calC,\CC)\oplus \CH^{1,0}(\calC,\CC)\oplus \CH^{0,1}(\calC,\CC)\oplus \CH^{1,1}(\calC,\CC)\ ,
\ee where 
\be \CH^{0,0}(\calC,\CC) =\CC \left|- \tfrac{1}{2},+ \tfrac{1}{2}\right\rangle\ ,\qquad \CH^{1,1}(\calC,\CC) =\CC \left|+ \tfrac{1}{2},- \tfrac{1}{2}\right\rangle
\ee span the vertical cohomology, and
\be \CH^{1,0}(\calC,\CC) =\CC \left|+ \tfrac{1}{2},+ \tfrac{1}{2}\right\rangle\ ,\qquad \CH^{0,1}(\calC,\CC) =\CC \left|- \tfrac{1}{2},- \tfrac{1}{2}\right\rangle
\ee span the middle cohomology.

The lattice of D-brane charges $\CH^\bullet(\calC,\ZZ)$ is the direct sum of the vertical and the middle sublattices
\be \CH^\bullet(\calC,\ZZ)=\CH^{vert}(\calC,\ZZ)\oplus \CH^{middle}(\calC,\ZZ)\ ,
\ee where the vertical contribution is spanned by charges of B-type D-branes ($D0$- and $D2$-branes, or BPS bound states thereof)
\begin{align} \CH^{vert}(\calC,\ZZ)=&\ZZ\left(\frac{|+-\rangle +i|-+\rangle}{2\sqrt{\rho_2}}\right)\oplus \ZZ\left(\frac{\rho|+-\rangle +i\bar\rho |-+\rangle}{2\sqrt{\rho_2}}\right)\\
=&\left\{\frac{1}{2\sqrt{\rho_2}}\begin{pmatrix}
    k_0 & k_2
\end{pmatrix}\begin{pmatrix}1 & 1\\\rho & \bar\rho  \end{pmatrix}\begin{pmatrix}|+-\rangle \\i|-+\rangle   \end{pmatrix}\mid k_0,k_2\in \ZZ\right\}
\end{align} while the middle contribution is spanned by charges of A-type D-branes ($D1$-branes)
\begin{align} \CH^{middle}(\calC,\ZZ)&=\ZZ\left(\frac{|++\rangle -i|--\rangle}{2\sqrt{\tau_2}}\right)\oplus \ZZ\left(\frac{\bar\tau|++\rangle -i\tau |--\rangle}{2\sqrt{\tau_2}}\right)\\
=&\left\{\frac{1}{2\sqrt{\tau_2}}\begin{pmatrix}
    k_{\alpha} & k_{\beta}
\end{pmatrix}\begin{pmatrix}1 & 1\\\bar\tau & \tau  \end{pmatrix}\begin{pmatrix}|++\rangle \\-i|--\rangle   \end{pmatrix}\mid k_{\alpha},k_{\beta}\in \ZZ\right\}\ .
\end{align}
These explicit formulae are obtained by projecting on $\CH^\bullet(\calC,\CC)$ the boundary states of a suitable set of BPS D-branes (see appendix \ref{a:Dbranes} for details).

Let us now consider the topological defects in this model. The $T^2$ sigma model always contains a group \be U(1)^4_{w-m}:=U(1)_{n_1}\times U(1)_{w_1}\times U(1)_{n_2}\times U(1)_{w_2}\ee of invertible topological defects $\CL_{a_k,b_k}$ acting on states carrying  winding-momentum by
\be \hat\CL_{a_k,b_k}|n_k,w_k\rangle= e^{2\pi i (a_1n_1+a_2n_2+b_1w_1+b_2w_2)}|n_k,w_k\rangle\ ,\qquad a_1,a_2,b_1,b_2\in \RR/\ZZ\ .
\ee These invertible symmetries generate the category of  defects that preserve the whole holomorphic  and anti-holomorphic algebra generated by the free fermions and free bosons. All such defects act trivially on the R-R ground states and therefore the D-brane charges -- they only shift the D-branes moduli.

We now want to discuss some more potential topological defects that preserve the $\CN=(2,2)$ superconformal algebra, but possibly act non-trivially on the other holomorphic and anti-holomorphic fields (see also \cite{Caldararu:2025eoj} for a detailed discussion about non-invertible symmetries on topological B-model on elliptic curves). We still want to require the action of simple defects on the spectral flow generators to be invertible, as in \eqref{ActOnSpecFlow1} and \eqref{ActOnSpecFlow2}. Explicitly, this means that, when such a simple defect $\CL$ moved across the insertion of one of the NS-NS spectral flow operators, they transform as
\be\label{T2ActOnSpecFlow1} \psi^\pm \mapsto g_L(\CL)^{\pm 1}\psi^{\pm}\ ,\qquad \tilde\psi^\pm \mapsto g_R(\CL)^{\pm 1}\tilde\psi^{\pm}\ ,\ee
for some $g_L(\CL),g_R(\CL)\in S^1\subset \CC^*$ complex numbers of modulus $1$. In order to preserve the $\CN=2$ supercurrents, the supersymmetric descendants of these fields must transform in the same way, so that
\be\label{T2ActOnSpecFlow2} i\partial Z \mapsto g_L(\CL)i\partial Z\ ,\qquad  i\partial Z^* \mapsto \overline{g_L(\CL)}i\partial Z^*\ee
\be\label{T2ActOnSpecFlow3} i\bar\partial Z \mapsto g_R(\CL)i\bar\partial Z\ ,\qquad  i\bar\partial Z^* \mapsto \overline{g_R(\CL)}i\bar\partial Z^*\ .\ee We denote by
\be U(1)_L^{\CN=2}\times U(1)_R^{\CN=2}
\ee the group of $(g_L(\CL),g_R(\CL))$ acting as in \eqref{T2ActOnSpecFlow1}--\eqref{T2ActOnSpecFlow3}; this is a group of outer automorphisms of $\CA\times\tilde\CA$ that leaves the $\CN=(2,2)$ algebra fixed.
The topological defects with these properties in a generic sigma model on $T^d$ were classified in \cite{Bachas:2012bj} (see also \cite{Bharadwaj:2024gpj}). Let us summarise their results in the case of $T^2$. 
Let us denote by $O(2,2,\ZZ)$ the group of automorphisms of the Narain lattice $\Gamma_{w-m}^{2,2}$ of winding-momenta. The corresponding groups of automorphisms of the rational and real vector spaces $\Gamma^{2,2}_{w-m}\otimes \QQ$ and $\Gamma^{2,2}_{w-m}\otimes \RR$ are $O(2,2,\QQ)$ and $O(2,2,\RR)$, respectively. The real Lie group $O(2,2,\RR)$ has four connected components, and the component connected to the identity is denoted by $SO^+(2,2,\RR)$. Similarly, one defines \be SO^+(2,2,\ZZ):=O(2,2,\ZZ)\cap SO^+(2,2,\RR)\qquad SO^+(2,2,\QQ):=O(2,2,\QQ)\cap SO^+(2,2,\RR)\ .\ee Notice that there is an isomorphism
\be SO^+(2,2,K)\cong (SL(2,K)_\tau\times SL(2,K)_\rho)/(-1,-1)\ ,
\ee where $K$ is either $\ZZ$, $\QQ$ or $\RR$. Explicitly, if $(n_1,w_1,n_2,w_2)$, with $n_i,w_i\in K$ and $K=\ZZ$, $\QQ$ or $\RR$, represent a vector in $\Gamma^{2,2}\otimes K$ of norm
\be 2(n_1w_1+n_2w_2)=2\det\begin{pmatrix}
    w_2 & w_1\\ -n_1 & n_2
\end{pmatrix}\ ,
\ee then, the element $(\gamma_\tau,\gamma_\rho)\in SL(2,K)_\tau\times SL(2,K)_\rho$ acts on this vector by
\be\label{transf} \begin{pmatrix}
    w_2 & w_1\\ -n_1 & n_2
\end{pmatrix}\mapsto \begin{pmatrix}
    w'_2 & w'_1\\ -n'_1 & n'_2
\end{pmatrix}=(\gamma_\rho^{-1})^t\begin{pmatrix}
    w_2 & w_1\\ -n_1 & n_2
\end{pmatrix}\gamma_\tau^{-1}\ .
\ee It is clear, using that $\det\gamma_\tau=1=\det\gamma_\rho$, that the norm of the vector is preserved.

When $K=\ZZ$, each $(\gamma_\tau,\gamma_\rho)\in SL(2,\ZZ)_\tau\times SL(2,\ZZ)_\rho$ also corresponds to a T-duality, establishing the equivalence, as SCFTs, of two sigma models on $T^2$ with moduli $(\tau,\rho)$ and $(\tau',\rho')$, where for $\gamma_\tau=\left(\begin{smallmatrix}
    a & b\\ c & d
\end{smallmatrix}\right)$ and $\gamma_\rho=\left(\begin{smallmatrix}
    \alpha & \beta\\ \gamma & \delta
\end{smallmatrix}\right)$ one has
\be \tau'=\frac{a\tau+b}{c\tau+d}\ ,\qquad \rho'=\frac{\alpha\rho+\beta}{\gamma\rho+\delta}\ .
\ee
Self-dual points $(\tau,\rho)$ in the moduli space are the points such that
\be\label{CMconds} \tau=\frac{a\tau+b}{c\tau+d}\ ,\qquad \rho=\frac{\alpha\rho+\beta}{\gamma\rho+\delta}\ .
\ee for some non-trivial\footnote{One obvious case, valid for any torus, is $(\gamma_\tau,\gamma_\rho)=(-1,1)$ or, equivalently, $(1,-1)$. This self-duality corresponds to the $\ZZ_2$ symmetry changing sign of all $Z$ and $\psi^\pm$, $\tilde \psi^{\pm}$, that is present for any torus sigma model. Here, by `non-trivial', we mean some different self-duality.}  $(\gamma_\tau,\gamma_\rho)\in SL(2,\ZZ)_\tau\times SL(2,\ZZ)_\rho$.  Solutions to this equations in the upper half-spaces only occur when either $\tau$ or $\rho$ (or both) are one of the elliptic points $i$ or $e^{\frac{2\pi i}{3}}$, or the image of one of these points via a $SL(2,\ZZ)$ transformation. In this case, the corresponding duality is just a (invertible) symmetry of the model $(\tau,\rho)$.

As discussed in \cite{Bachas:2012bj}, one can think of $SO^+(2,2,\QQ)\cong (SL(2,\QQ)_\rho\times SL(2,\QQ)_\tau)/(-1,-1)$ as an extension of the group of T-dualities, where each element $(\gamma_\tau,\gamma_\rho)$ corresponds (non uniquely) to a topological interface between the theory $(\tau,\rho)$ and the theory $(\tau',\rho')$. If $(\tau',\rho')=(\tau,\rho)$ for a given $(\gamma_\tau,\gamma_\rho)\in SO^+(2,2,\QQ)$, then the corresponding interface becomes a topological defect $\CL_{\gamma_\tau,\gamma_\rho}$ on the theory $(\tau,\rho)$. This happens when $(\tau,\rho)$ satisfy the conditions \eqref{CMconds} for some \emph{rational} (rather than integral) $(\gamma_\tau,\gamma_\rho)$. This means that non-trivial topological defects $\CL_{\gamma_\tau,\gamma_\rho}$ can only be present at very special points $(\tau,\rho)$ in the moduli space of $T^2$ sigma models, namely the ones where either $\tau$, or $\rho$, or both satisfy a quadratic equation
\be cx^2+(d-a)x-b=0\ ,
\ee for rational coefficients $a,b,c,d\in \QQ$. These points in the upper half-space are called Complex Multiplication (CM) points.

When a defect $\CL_{\gamma_\tau,\gamma_\rho}$ is moved across the insertion of a free fermion $\psi^\pm,\tilde \psi^\pm$ or free boson $i\partial Z,i\partial Z^*,i\bar\partial Z,i\bar\partial Z^*$, the action is as in \eqref{T2ActOnSpecFlow1}--\eqref{T2ActOnSpecFlow3}, with
\be g_L(\CL_{\gamma_\tau,\gamma_\rho})=\left(\gamma\bar \rho+\delta\right)\left(c\tau+d\right)\qquad g_R(\CL_{\gamma_\tau,\gamma_\rho})=\left(\gamma \rho+\delta\right)\left(c\tau+d\right)\ .
\ee where $\gamma_\tau=\left(\begin{smallmatrix}
    a & b\\ c & d
\end{smallmatrix}\right)$ and $\gamma_\rho=\left(\begin{smallmatrix}
    \alpha & \beta\\ \gamma & \delta
\end{smallmatrix}\right)$. One can show that, whenever \eqref{CMconds} is satisfied for some $(\gamma_\tau,\gamma_\rho)$, then $g_L(\CL_{\gamma_\tau,\gamma_\rho})$ and $g_R(\CL_{\gamma_\tau,\gamma_\rho})$ are complex number of unit modulus (see appendix \ref{a:units}). In fact, the pairs $\gamma_\tau$ and $\gamma_\rho$ in $SL(2,\QQ)$ for which \eqref{CMconds} are satisfied are in one-to-one correspondence with
\be \{c\tau+d\mid c,d\in \QQ,\ |c\tau+d|=1\}=(\QQ+\tau\QQ)\cap S^1\ ,
\ee and 
\be \{\gamma\rho+\delta \mid \gamma,\delta\in \QQ,\ |\gamma\rho+\delta|=1\}=(\QQ+\rho\QQ)\cap S^1\ ,
\ee i.e. points of modulus $1$ in the number fields $\QQ+\tau\QQ$ and $\QQ+\rho\QQ$. One can show that if $x=\tau$ or $\rho$ is a CM point, then the intersection $(\QQ+x\QQ)\cap S^1$ contains infinitely many distinct points (see appendix \ref{a:units}).

The action of $\CL_{\gamma_\tau, \gamma_\rho}$ on the states $|p,\tilde p\rangle$ with non-zero winding momentum is, in general, non-invertible. Indeed, the rational transformations $(\gamma_\tau,\gamma_\rho)$ do not map $\Gamma^{2,2}$ to itself, but to a lattice
\be (\gamma_\rho^{-1})^t\Gamma\gamma_\tau^{-1}=\left\{(\gamma_\rho^{-1})^t\begin{pmatrix}
    w_2 & w_1\\ -n_1 & n_2
\end{pmatrix}\gamma^{-1}_\tau\mid n_1,n_2,w_1,w_2\in \ZZ\right\}\subset \Gamma^{2,2}\otimes\QQ
\ee that is still even unimodular.
The action on the states $|p,\tilde p\rangle$ is
\be \hat\CL_{\gamma_\tau,\gamma_p}|p,\tilde p\rangle=\begin{cases}
    \langle \CL_{\gamma_\tau,\gamma_\rho}\rangle \xi_{\gamma_\tau,\gamma_\rho}(p,\tilde p)|p',\tilde p'\rangle & \text{if }
(p',\tilde p')\in \Gamma^{2,2}\cap (\gamma_\rho^{-1})^t\Gamma\gamma_\tau^{-1}\\
0 & \text{otherwise,}
\end{cases}
\ee where
\be p'=g_L(\CL_{\gamma_\tau,\gamma_\rho})p\ ,\qquad \tilde p'=g_R(\CL_{\gamma_\tau,\gamma_\rho})\tilde p\ ,
\ee $\langle \CL_{\gamma_\tau,\gamma_\rho}\rangle$ is the quantum dimension
and  $\xi_{\gamma_\tau,\gamma_\rho}(p,\tilde p)$ is a suitable phase, that is determined only up to multiplication by a $U(1)_{w-m}^4$ phase. In fact, for each pair $(\gamma_\tau,\gamma_\rho)\in SO^+(2,2,\QQ)$ satisfying \eqref{CMconds}, there are actually infinitely many topological defects, related to each other by fusion with invertible defects  in $U(1)_{w-m}^4$. 

One can verify that the quantum dimension $\langle \CL_{\gamma_\tau,\gamma_\rho}\rangle$ is always a positive integer. More precisely, one has $\langle \CL_{\gamma_\tau,\gamma_\rho}\rangle=\langle \CL_{\gamma_\tau,1}\rangle\langle \CL_{1,\gamma_\rho}\rangle$, where $\langle \CL_{\gamma_\tau,1}\rangle$ and $\langle \CL_{1,\gamma_\rho}\rangle$ are the smallest positive integers such that $\langle \CL_{\gamma_\tau,1}\rangle\gamma_\tau$ and $\langle \CL_{1,\gamma_\rho}\rangle\gamma_\rho$ are integral matrices. See appendix \ref{a:qdim} for a proof.


The action of $\hat\CL_{\gamma_\tau,\gamma_\rho}$ on the R-R ground fields is
\be \hat\CL_{\gamma_\tau,\gamma_\rho} |s,\tilde s\rangle=\langle \CL_{\gamma_\tau,\gamma_\rho} \rangle g_L(\CL_{\gamma_\tau,\gamma\rho})^sg_R(\CL_{\gamma_\tau,\gamma\rho})^{\tilde s} |s,\tilde s\rangle\ .
\ee Explicitly, the action on the components $\CH^{r,s}(\calC,\CC)$ is
\begin{align}
    \CH^{0,0}(\calC,\CC):&\qquad   &&\hat\CL_{\gamma_\tau,\gamma_\rho}|- \tfrac{1}{2},+ \tfrac{1}{2}\rangle=\langle \CL_{\gamma_\tau,\gamma_\rho} \rangle (\gamma\rho+\delta)|- \tfrac{1}{2},+ \tfrac{1}{2}\rangle\ ,\\
    \CH^{1,1}(\calC,\CC):&\qquad   &&\hat\CL_{\gamma_\tau,\gamma_\rho}|+\tfrac{1}{2},- \tfrac{1}{2}\rangle=\langle \CL_{\gamma_\tau,\gamma_\rho} \rangle (\gamma\bar\rho+\delta)|+ \tfrac{1}{2},- \tfrac{1}{2}\rangle\ ,\\
    \CH^{1,0}(\calC,\CC):&\qquad   &&\hat\CL_{\gamma_\tau,\gamma_\rho}|+\tfrac{1}{2},+ \tfrac{1}{2}\rangle=\langle \CL_{\gamma_\tau,\gamma_\rho} \rangle (c\tau+d)|+ \tfrac{1}{2},+ \tfrac{1}{2}\rangle\ ,\\
    \CH^{0,1}(\calC,\CC):&\qquad   &&\hat\CL_{\gamma_\tau,\gamma_\rho}|-\tfrac{1}{2},- \tfrac{1}{2}\rangle=\langle \CL_{\gamma_\tau,\gamma_\rho} \rangle (c\bar\tau+d)|- \tfrac{1}{2},- \tfrac{1}{2}\rangle\ .
\end{align}
The transformations on the B-type D-brane charges is
\begin{align}&
    \frac{1}{2\sqrt{\rho_2}}\begin{pmatrix}
    k_0 & k_2
\end{pmatrix}\begin{pmatrix}1 & 1\\\rho & \bar\rho  \end{pmatrix}\begin{pmatrix}\hat\CL_{\gamma_\tau,\gamma_\rho}|+-\rangle \\i\hat\CL_{\gamma_\tau,\gamma_\rho}|-+\rangle   \end{pmatrix}=\frac{\langle \CL_{\gamma_\tau,\gamma_\rho}\rangle}{2\sqrt{\rho_2}}\begin{pmatrix}
    k_0 & k_2
\end{pmatrix}\begin{pmatrix}\gamma\bar \rho+\delta & \gamma\rho+\delta\\\rho(\gamma\bar\rho+\delta) & \bar\rho(\gamma\rho+\delta)  \end{pmatrix}\begin{pmatrix}|+-\rangle \\i|-+\rangle\end{pmatrix}\notag\\
&=\frac{\langle \CL_{\gamma_\tau,\gamma_\rho}\rangle}{2\sqrt{\rho_2}}\begin{pmatrix}
    k_0 & k_2
\end{pmatrix}\begin{pmatrix}-\gamma \rho+2\gamma\Re\rho+\delta & -\gamma \bar\rho+2\gamma\Re\rho+\delta\\\gamma|\rho|^2+\delta\rho & \gamma|\rho|^2+\delta\bar\rho  \end{pmatrix}\begin{pmatrix}|+-\rangle \\i|-+\rangle\end{pmatrix}\\
&=\frac{\langle \CL_{\gamma_\tau,\gamma_\rho}\rangle}{2\sqrt{\rho_2}}\begin{pmatrix}
    k_0 & k_2
\end{pmatrix}\begin{pmatrix}\alpha & -\gamma\\-\beta & \delta  \end{pmatrix}\begin{pmatrix}1 & 1\\\rho & \bar\rho  \end{pmatrix}\begin{pmatrix}|+-\rangle \\i|-+\rangle\end{pmatrix}
\end{align} where we used that the relation \eqref{CMconds} implies $\alpha=\delta+2\gamma\Re\rho$ and $\beta=-\gamma|\rho|^2$.
Similarly, the transformation on the A-type D-brane charges is
\begin{align}
    \frac{1}{2\sqrt{\tau_2}}\begin{pmatrix}
    k_{\alpha} & k_{\beta}
\end{pmatrix}&\begin{pmatrix}1 & 1\\\bar\tau & \tau  \end{pmatrix}\begin{pmatrix}\hat\CL_{\gamma_\tau,\gamma_\rho}|++\rangle \\-i\hat\CL_{\gamma_\tau,\gamma_\rho}|--\rangle   \end{pmatrix}=\notag\\
&=\frac{\langle \CL_{\gamma_\tau,\gamma_\rho}\rangle}{2\sqrt{\tau_2}}\begin{pmatrix}
    k_{\alpha} & k_{\beta}
\end{pmatrix}\begin{pmatrix}a & -c\\-b & d  \end{pmatrix}\begin{pmatrix}1 & 1\\\bar\tau & \tau  \end{pmatrix}\begin{pmatrix}\hat\CL_{\gamma_\tau,\gamma_\rho}|++\rangle \\-i\hat\CL_{\gamma_\tau,\gamma_\rho}|--\rangle\end{pmatrix}\ .
\end{align} Thus, because the matrices
\be \langle \CL_{\gamma_\tau,\gamma_\rho}\rangle\begin{pmatrix}\alpha & -\gamma\\-\beta & \delta  \end{pmatrix}\ ,\qquad \langle \CL_{\gamma_\tau,\gamma_\rho}\rangle\begin{pmatrix}a & -c\\-b & d  \end{pmatrix}
\ee have integral entries, one obtains that the transformation $\hat\CL_{\gamma_\tau,\gamma_\rho}$ maps the lattice $\CH^\bullet(\calC,\ZZ)$ to itself, as expected, and therefore it is a Hodge endomorphism $\EndL_{\gamma_\tau,\gamma_\rho}$. 

Thus, the algebra generated by $\QQ$-linear combinations of $\EndL_{\gamma_\tau,\gamma_\rho}$ corresponding to defects $\CL_{\gamma_\tau,\gamma_\rho}$ is isomorphic to
\be K_\tau\times K_\rho\ ,
\ee where the field $K_x$, for a given element $x$ in the upper half-plane, is defined by
\be K_x:=\begin{cases}
    \QQ+x\QQ & \text{if $x$ is CM}\ ,\\
    \QQ & \text{if $x$ is not CM}\ ,
\end{cases}
\ee where an element $x\in \CC$, $\Im x>0$, is CM if it is the solution of a quadratic equation with rational coefficients.

The Hodge loci in the moduli space of sigma models on elliptic curves are now easily determined. Indeed, the algebra $K_\tau\times K_\rho$ is non-trivial if and only if at least one of the two moduli $\tau$ and $\rho$ is a CM point. Thus, a Hodge locus is given by a set of $(\tau,\rho)$ with either $\tau$ a fixed CM point and $\rho$ any point in the upper half-plane, or vice versa. 

Furthermore, the algebra $K_\tau\times K_\rho$ has maximal dimension over $\QQ$, i.e. $\dim_\QQ(K_\tau\times K_\rho)=4$, if and only if both $\tau$ and $\rho$ are CM points. We conclude that a supersymmetric sigma model $\calC$ on $T^2$ is a CM SCFT, according to our definition, if and only if both moduli $\tau$ and $\rho$ are CM points. Notice that these are isolated points in the moduli space.

Let us also recall that the torus sigma model  $\calC$ is rational if and only if it is a CM SCFT, and in addition the two CM fields are isomorphic $K_\tau\cong K_\rho$ (see for example \cite{Gukov:2002nw}). This is a strictly stronger condition than just being CM: for example, the torus model with $\tau=i$ and $\rho=e^{\frac{2\pi i}{3}}$ is a CM SCFT, but it is not rational.

\subsection{K3 surfaces}
\label{s:K3_surfaces}

In this section we analyze the class of Hodge loci in the moduli space of $2d$ non-linear-sigma models with target space a K3 surface where non-trivial Hodge endomorphisms, induced by topological defects in the category $\TDL$, emerge. We also describe the conditions under which a subset of these loci correspond to CFTs of complex multiplication type.

Let us begin with a brief discussion of the relation between the R-R ground state sector $\mathcal{H}^{\bullet}(\mathcal{C}, \CC)$ of these models and the complex cohomology $H^{\bullet}(K3,\mathbb{C})$ of the underlying $K3$ target space. Although the two are isomorphic as complex vector spaces, $\mathcal{H}^{\bullet}(\mathcal{C}, \CC) \cong H^{\bullet}(K3,\mathbb{C})$, their associated Hodge decompositions, and consequently the corresponding rational Hodge structures and polarizations, do not exactly coincide but they can be identified in a non-trivial way.
First of all, the moduli of the sigma model depend on the geometric data given by the complex structure and the K\"ahler form together with the closed $2$-form B-field, which are naturally unified within the framework of generalized geometry \cite{huybrechts:2004dhu}. A generalized $K3$ structure consists of a pair of two poly-forms $( \varphi, \varphi') \in \mathcal{A}^{2 \ast}_{\mathbb{C}}(X)$, on the space of even-degree complex forms of the $K3$ manifold $X$, satisfying the following conditions:
\begin{equation}
    \begin{split}
        & \bullet \, \, d \varphi \, = \, 0 \, \, \, , \, \, \, d \varphi' \, = \, 0 \, \\
        &\bullet \, \, \langle  \varphi , \varphi \rangle \, = \, 0 \, \, \, , \, \, \, \langle  \varphi' , \varphi' \rangle \, = \, 0 \, ,  \\
        & \bullet \, \, \langle \varphi , \overline{\varphi} \rangle \, > \, 0 \, \, \, , \, \, \, \langle \varphi' , \overline{\varphi}' \rangle \, > \, 0 \, ,\\
        & \bullet \, \, \langle \varphi , \varphi' \rangle \, = \, 0 \quad (\text{K\"ahler}) \, ,  \\
        & \bullet \, \, \langle  \varphi , \bar{\varphi} \rangle = \langle  \varphi' , \bar{\varphi}' \rangle \quad (\text{hyperk\"ahler}) \, , \\
    \end{split}
    \label{e:K3_generalized_structure}
\end{equation}
where 
\begin{equation}
    \langle \varphi , \psi \rangle \equiv \, \int_{X} \left[ - \varphi_0 \wedge \psi_4 + \varphi_2 \wedge \psi_2 - \varphi_4 \wedge \psi_0  \right] \, \, \in \, \, \mathbb{C} 
\end{equation}
is the Mukai paring between the pair of elements  $\varphi \, , \, \psi \, \in \mathcal{A}^{2 \ast}_{\mathbb{C}}$.
Such structure generalizes the notions of symplectic and complex structures on $X$, in the sense that both can be viewed as particular cases of generalized Calabi–Yau structures which are respectively $\varphi=e^{i \omega}$ and $\varphi'= \sigma$ with $\omega$ a standard symplectic structure in $X$ and $\sigma$ the unique (up to a phase) $(2,0)$-form normalized so that $\int_X \sigma \wedge \bar{\sigma}= Vol(X)$. All the other possible generalized Calabi-Yau structures can be obtained as the B-field transform of these two simple structures defined via the following exterior product:
\begin{equation}
    e^{B} \cdot \varphi = \left( 1+ B+ \frac{1}{2} B \wedge B \right) \wedge \varphi \, . 
\end{equation}

For each generalized Calabi-Yau structure $\varphi$, its real and imaginary parts span a real $2$-plane $P_{\varphi}$ in the vector space $\mathcal{A}^{2 \ast}_{\mathbb{C}}$. Analogously, the real and imaginary parts of the associated cohomology class $\left[ \varphi \right]$ identify a  $2$-plane $P_{\left[ \varphi \right]} \subset H^{\bullet} (X, \mathbb{R})$. The $2$-plane $P_{\varphi}$ (or $P_{\left[ \varphi \right]}$) comes with a natural orientation from the ordered basis $(Re \varphi , Im \varphi)$, and all its vectors are positive with respect to the Mukai paring. 
Given a pair of generalized structures $(\varphi, \varphi') \in \mathcal{A}^{2 \ast }_{\mathbb{C}}$, the hyperk\"ahler condition in \eqref{e:K3_generalized_structure} translates into the requirement that  $P_{\left[ \varphi \right]}$ and $P_{\left[\varphi' \right]}$ are orthogonal (K\"ahler condition) and  $\langle \varphi , \overline{\varphi} \rangle = \langle \varphi' , \overline{\varphi}' \rangle $. \\
 Thus, to any generalized $K3$ structure $(\varphi, \varphi')$ on $X$, we can associate a pointwise oriented $4$-plane $\Pi_{(\varphi, \varphi')} \in \mathcal{A}^{2 \ast} (M)$ spanned by the two $2$-planes $P_{\varphi}$ and $P_{\varphi'}$. In the same way we can define the oriented positive $4$-plane $\Pi_{\left( \left[ \varphi \right], \left[ \varphi'\right] \right)}$ spanned by $P_{\left[ \varphi \right]}$ and $P_{\left[ \varphi' \right]}$ inside the cohomology $H^{\bullet}(X, \mathbb{R})$. Importantly, each plane $\Pi_{(\varphi, \varphi')}$ is always a $B$-field transform of the geometric $4$-plane $\Pi_{(\sigma, e^{i \omega})}$. This means that the positive $4$-plane inside the real cohomology is the $B$-field transform of the 4-plane spanned by
\begin{equation}
    \Pi_{(\left[ \sigma \right], \left[ e^{i \omega} \right])} \, = \, \langle \, Re\, \sigma , \, Im\,\sigma, \, \omega , \, 1- \frac{1}{2} \omega^2 \, \rangle
    \label{e:Pi_geometric}
\end{equation}
where $\sigma$ and $\bar{\sigma}$ are the unique $(2,0)$- and $(0,2)$-forms, $\omega$ is the $(1,1)$ K\"ahler form and $1- \frac{1}{2} \omega^2$ is a combination of the $0$- and the $4$-form. \\
Treating the $(0,0)$- and the $(2,2)$-form on an equal footing with the $2$-forms, the rational structure on the complex cohomology $V_{\mathbb{C}}=H^{\bullet}(X,\mathbb{C})$ is induced by the $\QQ$-extension of the Mukai lattice:
\begin{equation}
    H^{\bullet} (X, \mathbb{Z}) \cong U^{\oplus 4} \oplus E_8^{\oplus 2}(-),
\end{equation}
where $U$ denotes the hyperbolic lattice
\begin{equation}
    U= \left( \begin{matrix} 0 & 1 \\ 1 & 0 \end{matrix} \right)
\end{equation}
of signature $(1,1)$, while $E_8$ is the rank-$8$ negative-definite lattice defined by the $E_8$ Cartan matrix. The resulting rational structure $\left( H^{\bullet}(X,\mathbb{Q}) , H^{\bullet}(X,\mathbb{C}) \right)$ has overall signature $(4,20)$, with positive directions identified by $\Pi_{\left( \left[ \varphi \right], \left[ \varphi' \right] \right)}$.

What has been described so far corresponds to the geometric side of the picture. Let us now turn to the physical perspective and consider the non-linear sigma model $\mathcal{C}$ with target space a K3 surface, which realizes the small $\mathcal{N}=(4,4)$ superconformal algebra with central charge $(c, \tilde{c})=(6,6)$. The small $\mathcal{N}=(4,4)$ superconformal algebra is obtained by extending the $\mathcal{N}=(2,2)$ algebra in \eqref{e:N=2SCA} by the NS-NS spectral flow generators $V_{(\pm1,0)}$ and $\tilde{V}_{(0,\pm1)}$ corresponding to the $su(2)_1$ currents $J^{\pm}$ and $\tilde{J}^{\pm}$. The generators of the holomorphic $\mathcal{N}=4$ algebra are the four supercurrents $G^{\pm}(z),\, G'^{\pm}(z)$ of conformal weight $(\frac{3}{2},0)$, the stress energy tensor $T(z)$ and the $su(2)_1$ Kac-Moody algebra with currents $J^3(z),J^{\pm}(z)$. The commutation relations are the following:
\begin{equation}
    \begin{split}
        \left[ L_m , L_n \right] =  (m-n) L_{m+n}  + & \frac{1}{2} (m^3-m) \delta_{m+n,0}\\
        \left[L_m,J^3_n \right] =-nJ^3_{m+n} \quad \quad \quad & \quad \quad \quad \left[L_m,J^{\pm}_n \right] =-nJ^{\pm}_{m+n} \\
        \left[ 2J^3_m,2J^3_n\right]= 2 m \delta_{m+n,0} \quad \quad \quad & \quad \quad \quad \left[ J^{\pm} , J^{\pm} \right] =0 \\
        \left[ J^3_m,J^{\pm}_n\right]= \pm J^{\pm}_{m+n} \quad \quad \quad & \quad \quad \quad \left[ J^+_m, J^-_n \right]=m \delta_{m+n,0}+2J^3_{m+n} \\
        \left[ L_m, G_r^{\pm} \right] = \left(  \frac{m}{2} -r\right) G^{\pm}_{m+r} \quad \quad \quad & \quad \quad \quad  \left[ L_m, {G'}_r^{\pm} \right] = \left(  \frac{m}{2} -r\right) {G'}^{\pm}_{m+r}  \\
        \left[ J^3_m,G^{\pm}_r\right]= \pm \frac{1}{2} G^{\pm}_{m+r} \quad \quad \quad & \quad \quad \quad \left[ J^3_m,G'^{\pm}_r\right]= \mp \frac{1}{2} {G'}^{\pm}_{m+r} \\
        \left[ J^{\pm}_m,G^{\mp}_r\right]= \pm \frac{1}{2} {G'}^{\mp}_{m+r} \quad \quad \quad & \quad \quad \quad \left[ J^{\pm}_m,G^{\pm}_r\right]=0 \\
        \left[ J^{\pm}_m,{G'}^{\pm}_r\right]= \mp \frac{1}{2} G^{\pm}_{m+r} \quad \quad \quad & \quad \quad \quad \left[ J^{\pm}_m,{G'}^{\mp}_r\right]=0 \\
        \left\lbrace G^{\pm}_r , {G'}^{\mp}_s \right\rbrace =2(s-r)J^{\pm}_{s+r} \quad \quad \quad & \quad \quad \quad \left\lbrace G^{\pm}_r , {G'}^{\pm}_s \right\rbrace =0 \\
        \left\lbrace G_r^+,G_s^-\right\rbrace = \left\lbrace {G'}^+_r,{G'}^-_s\right\rbrace  = 2L_{r+s}& \pm  2(r-s)J^3_{r+s} +2\left( r^2 - \frac{1}{4} \right) \delta_{r+s,0} \\
    \end{split} 
\end{equation}
where $r, \, s \in \frac{1}{2}+ \mathbb{Z}$ in the NS sector and $r, \, s \in \mathbb{Z}$ in the R sector.\\
This algebra contains a whole family of $\mathcal{N}=(2,2)$ subalgebras parameterized by $S^2 \times S^2$, where each $S^2$ corresponds to the choice of a $\mathcal{N}=1$ subalgebra and a $u(1)$ inside the $su(2)$ R-symmetry algebra in the holomorphic and anti-holomorphic sectors. 
The spectrum of any $K3$ model contains exactly $24$ R-R ground states transforming in the massless unitary representations of the $\mathcal{N}=(4,4)$ algebra. They consists of a single $(\mathbf{2}, \mathbf{2})$ multiplet with quantum numbers $\left(h, q_{J^3_0} ; \tilde{h}, \tilde{q}_{\tilde{J}^3_0} \right)= \left( \frac{1}{4},  \frac{1}{2} ; \frac{1}{4} ,  \frac{1}{2} \right)$, which corresponding operators are the four R-R spectral flow generators, together with $20$ singlets labeled by $\left(h, q_{J^3_0} ; \tilde{h}, \tilde{q}_{\tilde{J}^3_0} \right)=\left( \frac{1}{4},  0 ; \frac{1}{4} ,  0 \right)$. \\
Once a specific $\mathcal{N}=(2,2)$ subalgebra is chosen inside the $\mathcal{N}=(4,4)$ algebra by fixing the $u(1)$ currents $j$ and $ \tilde{j}$, one can endow the $24$-dimensional space $\mathcal{H}(\mathcal{C}, \mathbb{C})$ of R-R ground states with an $(r,s)$-grading determined by the corresponding $U(1)$ R-charges, as defined in \eqref{e:grading_GS_space}.  In particular, using the rescaling $j_0=2J^3_0$ and $\tilde{j}_0=2 \tilde{J}^3_0$, the $24$ R-R ground states are now labeled by the quantum numbers $\left(h, q ; \tilde{h}, \tilde{q} \right)= \left( \frac{1}{4}, \pm 1 ; \frac{1}{4} , \pm 1 \right)$ and $\left(h, q ; \tilde{h}, \tilde{q} \right)= \left( \frac{1}{4}, 0 ; \frac{1}{4} , 0 \right)$. Thus, the space of R-R ground states decomposes as
\begin{equation}
    \mathcal{H}(\mathcal{C}, \mathbb{C}) \, = \,  \mathcal{H}^{2,2}(\mathcal{C}, \mathbb{C}) \, \oplus \, \mathcal{H}^{2,0}(\mathcal{C}, \mathbb{C}) \, \oplus \, \mathcal{H}^{1,1} (\mathcal{C}, \mathbb{C}) \, \oplus \, \mathcal{H}^{0,2} (\mathcal{C}, \mathbb{C}) \, \oplus \, \mathcal{H}^{0,0}(\mathcal{C}, \mathbb{C}),
\end{equation}
where 
\begin{equation}
    \begin{split}
        & \mathcal{H}^{2,2}(\mathcal{C}, \mathbb{C}) \, \oplus \, \mathcal{H}^{2,0}(\mathcal{C}, \mathbb{C}) \, \oplus \, \mathcal{H}^{0,2} (\mathcal{C}, \mathbb{C}) \, \oplus \, \mathcal{H}^{0,0}(\mathcal{C}, \mathbb{C}) = \mathbb{C} \, \big\vert \frac{1}{4}, \pm \frac{1}{2} ; \frac{1}{4}, \pm \frac{1}{2} \big\rangle \\
        & \mathcal{H}^{1,1} (\mathcal{C}, \mathbb{C}) \, = \, \bigoplus_{i=1}^{20} \mathbb{C} \,  \big\vert \frac{1}{4}, 0; \frac{1}{4}, 0 \big\rangle_i
    \end{split}
\end{equation}
and Hodge vector $(h^{2,2},h^{2,0} , h^{1,1}, h^{0,2}, h^{0,0})=(1,1,20,1,1)$. 

As explained in the general construction of Section \ref{s:proposal}, in order to endow the complex vector space $\mathcal{H}(\mathcal{C},\mathbb{C})$ with a rational structure, we use $BPS$ boundary states. These can be defined in the closed string sector as solutions $\vert \mathcal{B} \rangle$ to the gluing conditions:
\begin{equation}
\begin{split}
    & \left( L_{n} -  \Omega ( \tilde{L}_{-n} ) \right) \vert \mathcal{B} \rangle = 0 \quad \quad \forall \, \, m \, \in \, \mathbb{Z} \\
    & \left( J_{m}^i -  \Omega ( \tilde{J}^j_{-m} ) \right) \vert \mathcal{B} \rangle = 0 \quad \quad \forall \, \, m \, \in \, \mathbb{Z} \\
    & \left( G^{I}_{r} - \Omega (\tilde{G}^{K}_{-r})  \right)\vert \mathcal{B} \rangle = 0  \quad \quad \quad \quad \quad \quad \quad  \forall \, \, r \, \in \, \mathbb{Z}/2 \\
\end{split}
\label{e:gluing_condition_(4,4)}
\end{equation}
where $I,K$ label the four surpercurrents in the holomorphic and anti-holomorphic sectors, while  $\Omega$ denotes an automorphism of the full $\mathcal{N}=(4,4)$ superconformal algebra
\begin{equation}
    \Omega \, \, \, : \, \,\, sVir_{\mathcal{N}=(4,4)} \, \, \mapsto \, \, sVir_{\mathcal{N}=(4,4)} 
\end{equation}
The automorphism group of the (anti-)holomorphic $\mathcal{N}=4$ algebra is $SU(2)_{inn} \times SU(2)_{out}$, under which the four supercurrents transform in the $(\mathbf{2}, \mathbf{2})$ representation. Here, $SU(2)_{inn}$ is generated by the zero modes of $J^3,J^{\pm}$, while $SU(2)_{out}$ does not, in general, extend to a symmetry of the full CFT. Since we are interested in boundary states preserving an $\mathcal{N}=1$ supercurrent $G$, specified by a linear combination of the four supercurrents $G^{\pm}, G'^{\pm}$ (and similarly $\tilde{G}^{\pm}, \tilde{G}'^{\pm}$ in the anti-holomorphic sector), the gluing automorphism $\Omega$ must be restricted to the diagonal subgroup $SO(3) \subset SU(2)_{inn} \times SU(2)_{out}$ fixing $G$ \cite{Ooguri:1996oyz} . \\ The ground states components of the R-R BPS boundary states of the theory span a $24$-dimensional lattice $\Gamma^{4,20} \equiv \mathcal{H}^{\bullet}(\mathcal{C}, \mathbb{Z})$ as defined in \eqref{e:BS_lattice}. Its rational extension $\mathcal{H}^{\bullet}(\mathcal{C}, \mathbb{Q})= \mathcal{H}^{\bullet}(\mathcal{C}, \mathbb{Z}) \otimes \mathbb{Q}$ induces a rational structure $\left( \mathcal{H}^{\bullet}(\mathcal{C}, \mathbb{Q}), \mathcal{H}^{\bullet}(\mathcal{C}, \mathbb{C}) \right)$ on $V_{\mathbb{C}}=\mathcal{H}(\mathcal{C}, \mathbb{C})$. \\
The polarization on $\mathcal{H}(\mathcal{C}, \mathbb{C})$ is induced by the $\mathbb{C}$-linear extension of the open string Witten index $Q(w_{\mathcal{B}}, w_{\mathcal{B}'})$ between two boundary states $\mathcal{B}$ and $\mathcal{B}'$ whose R-R ground states components are $w_{\mathcal{B}},\, w_{\mathcal{B'}} \in \mathcal{H}(\mathcal{C}, \mathbb{Z})$, as defined in \eqref{e:Witten_index}. The vector space $\mathcal{H}(\mathcal{C}, \mathbb{C})$, equipped with this pairing, has signature $(4,20)$. The associate positive-definite $4$-plane $\Pi_{\mathcal{C}} \subset \mathcal{H}(\mathcal{C}, \mathbb{C})$ is spanned by the four R-R ground states transforming in the representations $ \left( \frac{1}{4}, \frac{1}{2} ; \frac{1}{4} , \frac{1}{2}\right)$ of the $\mathcal{N}=(4,4)$ algebra and filling the graded subspaces $\mathcal{H}^{4} \oplus \CH^{2,0} \oplus \CH^{0,2} \oplus \CH^{0} \subset \CH(\mathcal{C}, \mathbb{C})$. Notice that, the Hodge grading of the CFT does not coincide with the geometric one. In particular, in the first case the plane $\Pi_{\mathcal{C}}$ is spanned by the components of degree $(2,0), (0,2), (0,0),(2,2)$, while in the geometric case the plane $\Pi$ is a B-field transform of \eqref{e:Pi_geometric}. However, there exists a lattice isomorphism $H^{\bullet} \left( X, \mathbb{Z}\right) \cong \mathcal{H} \left( \mathcal{C}, \mathbb{Z} \right)$ such that the $4$-plane $\Pi$ of generalized geometry is mapped exactly to the $4$-plane $\Pi_{\mathcal{C}}$ and the Mukai pairing corresponds to the Witten index. \\
In our general proposal, we construct boundary states using A-type and B-type gluing conditions for a chosen $\mathcal{N}=(2,2)$ superconformal subalgebra. In the present case, however, there is an entire $S^2 \times S^2$ family of possible choices of such a subalgebra inside $\mathcal{N}=(4,4)$, and different choices do not necessarily lead to the same decomposition into A-type and B-type branes. For this reason, for the moment we do not fix any specific $\mathcal{N}=(2,2)$ subalgebra and instead consider the lattice of boundary states constructed using the gluing conditions \eqref{e:gluing_condition_(4,4)}. \\
Given the choice of the gluing conditions above, it is natural to consider topological defects $\mathcal{L}$ acting invertibly on the $\mathcal{N}=(4,4)$ superconformal algebra by an $SO(3)_L \times SO(3)_R$ automorphism\footnote{The group acting on the R-R sector is actually the central extension $\left( SU(2)_L \times SU(2)_R \right)/(-1,-1)$.} $\left( g_L(\mathcal{L}), g_R(\mathcal{L}) \right)$.
The boundary states of the model furnish a module for the category $\TDL_{(4,4)}$ generated by such defects. This implies that every $\mathcal{L} \in \TDL_{(4,4)}$ induces an endomorphism $\mathsf{L}:\CH^{\bullet}(\mathcal{C}, \mathbb{Z}) \mapsto \CH^{\bullet}(\mathcal{C}, \mathbb{Z})$ on the lattice of R-R D-brane charges, which $\mathbb{R}$-extension $\mathsf{L}_{\mathbb{R}}$ preserves the decomposition $\mathcal{H}(\mathcal{C}, \mathbb{R}) = \Pi_{\mathcal{C}} \oplus \Pi_{\mathcal{C}}^{\perp}$. In other words, the matrix representation of $\mathsf{L}_{\RR}$, in a basis compatible with the above decomposition, is block-diagonal 
\begin{equation}
    L_{\mathbb{R}} \,  = \, \left( \begin{matrix}
        b_{4 \times 4} & 0 \\ 0 & b_{20 \times 20}
    \end{matrix} \right) \, ,
\end{equation}
and such that the restriction $\mathsf{L}_{\vert \Gamma^{4,20}}$ provides an endomorphism of the lattice.
The category $Top_{\Pi}$ studied in \cite{Angius:2024evd} (see also \cite{Angius:2024evd, Angius:2025zlm, Angius:2025ium, Angius:2025dcu} for further examples) is actually the subcategory of defects in $\TDL_{(4,4)}$ such that the block $b_{4 \times 4}$ is the identity times the quantum dimension. We stress that neither the category $\TDL_{(4,4)}$, nor the set of boundary states, depend on the choice of a $\mathcal{N}=(2,2)$ subalgebra. Let us now fix a $\mathcal{N}=(2,2)$ and let $SO(2)_L \times SO(2)_R \subset SO(3)_L \times SO(3)_R$ be the corresponding R-symmetry group. The associated category $\TDL$, as defined in the proposal, corresponds to the subcategory of defects $\mathcal{L} \in \TDL_{(4,4)}$ such that $\left( g_L(\mathcal{L}), g_R(\mathcal{L}) \right) \in SO(2)_L \times SO(2)_R$. Their matrix representation reduce to 
\begin{equation}
    L_{\mathbb{R}} \, = \, \left( \begin{matrix}
        \begin{matrix} d  \cos \theta_1 & -d \sin \theta_1 & 0 & 0 \\ d \sin \theta_1 & d \cos \theta_1 & 0 & 0 \\ 0 & 0 & d \cos \theta_2 & - d \sin \theta_2 \\ 0 & 0 & d \sin \theta_2 & d \cos \theta_2 \end{matrix} & \mathbf{0} \\ \mathbf{0} & b_{20 \times 20}
    \end{matrix} \right)\, .
\end{equation}
where $d= \langle \mathcal{L} \rangle \geq 1$. Here, the $2\times 2$ blocks correspond to the subspaces $\Pi_\calC=P_{[\varphi]}\oplus P_{[\varphi']}$ where $(\varphi,\varphi')$ define the generalized K3 structure. Whenever the category $\TDL$ contains at least one non-trivial defect, which is not present at generic points, the corresponding CFT model $\mathcal{C}$ lies on a Hodge locus of the $\mathcal{N}=(2,2)$ moduli space.\\
Suppose now  there is a subcategory $\TDL' \subset \TDL_{(4,4)}$ such that the corresponding algebra of lattice endomorphisms is isomorphic to a product of CM fields. This implies that the algebra is commutative, so that all the corresponding automorphisms $\left( g_L(\mathcal{L}), g_R(\mathcal{L}) \right)$ lie in an abelian subgroup of $SO(3)_L \times SO(3)_R$. Since all such abelian subgroups are contained in some $SO(2)_L \times SO(2)_R \subset SO(3)_L \times SO(3)_R $, we can conclude that $\TDL'$ fixes a $\mathcal{N}=(2,2)$ subalgebra. In addition, if $\TDL'$ contains enough defects to satisfy the main Definition \ref{d:CM_cft} in Section \ref{s:proposal}, the corresponding CFT $\mathcal{C}$ has CM with respect to the preserved $\mathcal{N}=(2,2)$. 

\begin{definition}
\label{d:CMK3}
A K3 sigma model $\calC$ is a Complex Multiplication SCFT if and only if there exists an orthogonal decomposition $\Hh^\bullet(\calC,\QQ)=\oplus_\alpha W^\alpha$ and a tensor subcategory $\TDL'$ of $\TDL_{(4,4)}$ such that the induced algebra of endomorphisms $\End_{\TDL'}$ is isomorphic to a product of CM fields $\prod_\alpha K_\alpha$ where $K_\alpha \subseteq \End_\QQ(W^\alpha)$ and $\dim_\QQ K_\alpha=\dim_\QQ W^\alpha$.
\end{definition}
By the argument above, $\End_{\TDL'}$ acts on $\Hh^\bullet(\calC,\QQ)$ by Hodge endomorphisms, where the grading is given by the particular $\CN=(2,2)$ subalgebra that is preserved by $\TDL'$. Therefore, the $W^\alpha$ are necessarily Hodge substructures.
Notice that, with respect to the general definition in section \ref{s:proposal}, we do not require the $W^\alpha$ to be irreducible.

Let us comment on the motivation for this different definition and its consequences.
In the pure geometric case of a K3 manifold $X$ equipped with a complex structure $\sigma$ and a symplectic K\"ahler form $\omega$, the rational Hodge structure is defined inside the middle cohomology $\left( H^2(X,\mathbb{Q}), H^2(X, \mathbb{C})\right)$ by the Mukai lattice $U^{\oplus 3} \oplus E_8^{\oplus 2}(-)$ and at some special points in the moduli space it can be reducible under the decomposition 
\begin{equation}
    H^2 \left( X, \mathbb{Q} \right) \, = \, Pic(X) \, \oplus \, T(X).
\end{equation}
where $Pic(X) \equiv H^{1,1}(X, \mathbb{C}) \cap H^2(X,\mathbb{Z})$ is the Picard lattice and $T(X)$ is the transcendental lattice. Importantly, if $X$ is an algebraic surface, then $T(X) \subset H^2(X, \mathbb{Q})$ defines an irreducible sub-Hodge structure of weight $2$ in $\left( H^2(X,\mathbb{Q}), H^2(X, \mathbb{C})\right)$ polarized by the Mukai pairing \cite{Huybrechts:2009COMPLEXAR}, while $Pic(X)$ defines a direct sum of one-dimensional pure Hodge structures of type $(1,1)$ which are irreducible. In this pure geometric setting the study of CM type reduces to the study of Hodge endomorphisms $\End_{Hdg}(T)$ of $T(X)$. In particular, for algebraic $K3$ surfaces, the algebra $\End_{Hdg}(T)$ is always commutative and isomorphic to a number field $K$ which can only be totally real or of complex multiplication type \cite{Huybrechts:2009COMPLEXAR}. If the second option is realized, the geometric $K3$ manifold is said to admit complex multiplication. \\
In our proposal to define a sigma model version of Hodge endomorphisms and Hodge structures of complex multiplication type we need to study reducibility of the rational Hodge structure $\left( \CH(\mathcal{C}, \mathbb{Q}), \CH(\mathcal{C}, \mathbb{C})\right)$ which involve both the middle and the vertical cohomologies. In the geometric language they can be seen as the middle cohomology of the target Calabi-Yau and the middle cohomology of the mirror Calabi-Yau (the vertical). In the CFT language with $\mathcal{N}=(2,2)$, the middle cohomology is spanned by R-R ground fields with R-symmetry charges $q= \tilde{q}$, while the vertical one by R-R ground fields with $q=-\tilde{q}$. In the case of a Calabi-Yau manifold $X$ of odd complex dimension, such as an elliptic curve or a $CY_3$, having CM type in the CFT sense corresponds to requiring that both $X$ and its mirror dual $\hat{X}$ are of CM type. In the case of Calabi-Yau manifolds with even complex dimension the two cohomologies intersect in $H^{d/2,d/2}$, and even at the level of the CFT there can be R-R ground fields that are simultaneously in $\CH^{middle}(\mathcal{C}, \CC)$ and $\CH^{vert}(\mathcal{C}, \CC)$, then there is not a clear splitting among the two. This is what happens for example to R-R ground fields transforming in the representations $\left( \frac{1}{4}, \frac{1}{2}; \frac{1}{4}, \frac{1}{2}\right)$ of the $\mathcal{N}=(4,4)$ algebra in K3 models. This non-trivial intersection prevents to naively translate the concept of CM in the CFT language with the requirement to have geometric CM for the $K3$ manifold $X$ and its mirror $\hat{X}$. This drawback is related to the subtle problem to define mirror symmetry for K3 surfaces \cite{Dolgachev} and for the corresponding sigma model \cite{Aspinwall:1996mn,Nahm:2001kh}. A second important point that distinguishes the geometric definition of CM type for $K3$ surfaces from our CFT version concerns the constraints on the rational $1$-dimensional substructures, as the one coming from the $Pic(X)$ lattice in the pure geometric case.

Let us now discuss how our definition \ref{d:CMK3} deals with this ambiguity. Suppose that $\calC$ is a CM SCFT, according to definition \ref{d:CMK3}. A generic R-R ground state $v\in \Hh^\bullet(\calC,\CC)=\bigoplus_\alpha W^\alpha_\CC$ decomposes as
\be v=\sum_\alpha v_\alpha\ ,
\ee with $v_\alpha \in W^\alpha_\CC$, and an endomorphism $\EndL\in \End_{\TDL'}$ acts by
\be\label{act1} \hat\CL v\equiv \EndL v=\sum_\alpha L_\alpha v_\alpha
\ee with $L_\alpha\in K_\alpha$. On the other hand, we know that every simple $\CL\in \TDL'$ acts on the charged R-R ground states by
\be \hat\CL \big\vert \frac{1}{4}, \pm \frac{1}{2} ; \frac{1}{4}, \pm \frac{1}{2} \big\rangle=g_L(\CL)^{\pm 1}g_R(\CL)^{\pm 1}\langle \CL\rangle \big\vert \frac{1}{4}, \pm \frac{1}{2} ; \frac{1}{4}, \pm \frac{1}{2} \big\rangle\ .
\ee More generally, any $\EndL\in \End_{\TDL'}$ acts on these states by
\be\label{act2} \EndL \big\vert \frac{1}{4}, \pm \frac{1}{2} ; \frac{1}{4}, \pm \frac{1}{2} \big\rangle=\EndL^{\pm\pm}\big\vert \frac{1}{4}, \pm \frac{1}{2} ; \frac{1}{4}, \pm \frac{1}{2} \big\rangle\ .
\ee where $\EndL^{\pm\pm}\in \CC$ are complex numbers such that $\EndL^{--}=\overline{\EndL^{++}}$ and $\EndL^{-+}=\overline{\EndL^{+-}}$. Compatibility of \eqref{act1} and \eqref{act2} for all 
$\EndL\in \End_{\TDL'}\cong \prod_\alpha K_\alpha$ implies that $\big\vert \frac{1}{4}, + \frac{1}{2} ; \frac{1}{4}, + \frac{1}{2} \big\rangle$ and $\big\vert \frac{1}{4}, - \frac{1}{2} ; \frac{1}{4}, - \frac{1}{2} \big\rangle$ are both contained in one $W^\alpha_\CC$ substructure; we denote such substructure by $W^B_\CC$. Similarly, $\big\vert \frac{1}{4}, + \frac{1}{2} ; \frac{1}{4}, - \frac{1}{2} \big\rangle$ and $\big\vert \frac{1}{4}, - \frac{1}{2} ; \frac{1}{4}, + \frac{1}{2} \big\rangle$ are both contained in a single $W^{\alpha'}_\CC$ with $\alpha'\neq \alpha$, and we denote such substructure by $W^A_\CC$. All the other substructures $W^\alpha_\QQ$, if present, must be contained in $H^{1,1}(\calC,\CC)\cap H^\bullet(\calC,\QQ)$. The R-R charge vector of a generic BPS boundary state has a non-zero component along all $W^\alpha$. However, we expect to have BPS boundary states that generate the sublattice $(W^B)^\perp \cap \Gamma^{4,20}$, that has rank $24-\dim_\QQ W^B_\QQ$. These D-branes are A-type D-branes with respect to the $\CN=(2,2)$ algebra fixed by the subcategory $\TDL'$. Analogously, the BPS boundary states with charge vector in the sublattice $(W^A_\QQ)^\perp \cap \Gamma^{4,20}$ correspond to B-type D-branes. 


Even in the case where $\calC$ is not a CM SCFT, this discussion suggests a natural way to address the splitting between the middle and vertical cohomologies and to identify the irreducible substructures of $\left( \CH(\mathcal{C}, \mathbb{Q}), \CH(\mathcal{C}, \mathbb{C})\right)$. We propose a democratic treatment, in close analogy with the democratic treatment of the Néron-Severi and transcendental lattices appearing in a recent formulation of mirror symmetry for generalized $K3$ surfaces proposed in \cite{Kanazawa:2025kaa}, which we briefly review below.\\ The central idea is to introduce generalized notions of the Néron-Severi and Transcendental lattices that treat the two generalized structures in the pair $(\varphi,\varphi')$ on equal footing. More precisely, given a generalized $K3$ surface $X$ equipped with a pair $(\varphi,\varphi')$, the generalized Néron-Severi and Transcendental lattices are defined respectively by
\begin{equation}
    \tilde{NS} (X) \, \equiv \, \left\lbrace \delta \in H^{\bullet} (X, \mathbb{Z}) \vert \langle \delta, \varphi' \rangle =0 \right\rbrace \, ,
    \label{e:genereralized_NS}
\end{equation}
\begin{equation}
    \tilde{T}(X) \, \equiv \, \left\lbrace \delta \in H^{\bullet} (X, \mathbb{Z}) \vert \langle \delta , \varphi \rangle =0\right\rbrace.
    \label{e:generalized_T}
\end{equation}
The price to pay for treating the two generalized structures $\varphi$ and $\varphi'$ democratically is that the lattices $\widetilde{NS}(X)$ and $\widetilde{T}(X)$ may now have a non-trivial intersection, $\tilde{NS} (X) \cap \tilde{T}(X) \, \neq \, \emptyset$. This is the analogous to have $H^{middle} \left( \mathcal{C}, \mathbb{Z}\right) \cap H^{vert} \left( \mathcal{C}, \mathbb{Z}\right) \neq \emptyset$ in the CFT formulation. The next step is to consider a pair $(K,L)$ of even lattices of signatures $(2,\kappa-2)$ and $(2,\lambda-2)$ respectively, satisfying $\kappa+\lambda=24$. One may then define a lattice-polarized generalized $K3$ surface as follows.

\vspace{0.15cm}
\noindent
\textbf{Definition:}[Lattice polarized generalized K3]\\
\textit{A $(K,L)$-polarized generalized K3 surface is a pair $(X,i)$ of a generalized K3 surface and a lattice embedding}
\begin{equation}
    i\, : \, K \oplus L \, \, \hookrightarrow \, \, H^{\bullet}(X, \mathbb{Z})
\end{equation}
\textit{such that \begin{itemize}
    \item[\textit{(i)}] the restrictions $i\vert_K$ and $i \vert_L$ are primitive embeddings;
    \item[\textit{(ii)}] $i(K) \subset \tilde{NS}(X)$ and $i(L) \subset \tilde{T}(X)$. 
\end{itemize}}
In terms of the previous structures, mirror symmetry for generalized $K3$ surfaces may be formulated as follows. 

\vspace{0.15cm}
\noindent
\textbf{Definition:}[Generalized Mirror Symmetry]\\
\textit{A family $\chi$ of $(K,L)$-polarized generalized K3 surfaces and a family $\Upsilon$ of $(L,K)$-polarized generalized K3 surfaces are mirror symmetric.}

In the CFT formulation, the two generalized structures $(\varphi,\varphi')$ are replaced by the choice of two two-planes $\Pi_A$ and $\Pi_B$, which decompose the four-plane $\Pi_{\mathcal{C}}$ as follows:
\begin{equation}
    \Pi_{\mathcal{C}} \, = \, \Pi_A \oplus \Pi_B.
\end{equation}
The choice of such a splitting is equivalent to selecting a specific embedding of $\mathcal{N}=(2,2)$ into $\mathcal{N}=(4,4)$. Moreover, we require the two 2-planes $\Pi_A$ and $\Pi_B$ to satisfy the condition that, for each of them, there exist at least two BPS boundary states constructed using the gluing conditions \eqref{e:gluing_condition_(4,4)} whose corresponding charge vectors, projected onto $\Pi_{\mathcal{C}}$, lie entirely within $\Pi_A$ and $\Pi_B$, respectively. Fixed this setting, in analogy with the definitions in \eqref{e:genereralized_NS} and \eqref{e:generalized_T}, we define the following two lattices
\begin{equation}
    \Gamma_A \, \equiv \, \left\lbrace \delta \in \Gamma^{4,20} \vert Q( \delta,  \Pi_B  )=0 \right\rbrace
\end{equation}
\begin{equation}
    \Gamma_B \, \equiv \, \left\lbrace \delta \in \Gamma^{4,20} \vert Q(\delta,  \Pi_A  )=0\right\rbrace
\end{equation}
where $Q(\,,\,)$ is the polarization induced by the Witten index. As in the geometric\footnote{Notice that with the term geometry, here we mean "generalized geometry".} case, in general we can have $\Gamma_A \cap \Gamma_B \neq \emptyset$. Notice that, for the particular choice $\Pi_A = \CH^{2,0} \oplus \CH^{0,2}$ and $\Pi_B= \CH^{2,2} \oplus \CH^{0,0}$, we have $\Gamma_A \otimes \mathbb{R} \subseteq \CH^{middle}(\mathcal{C}_{\Pi}, \mathbb{R})$ and $\Gamma_B \otimes \mathbb{R} \subseteq \CH^{vert}(\mathcal{C}_{\Pi}, \mathbb{R})$. \\
Now, we define the two mutually orthogonal sublattices 
\begin{equation}
    \begin{split}
        & T_A \, \equiv \, \Gamma_A \cap \Gamma_B^{\perp} \subseteq \Gamma_A \\
        & T_B \, \equiv \,  \Gamma_A^\perp \cap \Gamma_B  \subseteq \Gamma_B \\
    \end{split}
\end{equation}
which define by construction two irreducible Hodge substructure of $\left( H (\mathcal{C}, \mathbb{Q}), H (\mathcal{C}, \mathbb{C}) \right)$, while $\Gamma_A \cap \Gamma_B$ defines a direct sum of pure one-dimensional substructures. 
When $\calC$ has CM, then $T^A\otimes \QQ=W^A$ and $T^B\otimes \QQ=W^B$ are CM Hodge substructures.

\section{Concluding remarks and Outlook}
\label{s:conclusions}

In this article, we propose a sigma-model analogue of Hodge loci arising from the appearance of non-trivial Hodge tensors in the moduli space of geometric Calabi–Yau compactifications. Any $\mathcal{N}=(2,2)$ SCFT endowed with boundary states satisfying the standard A-type and B-type boundary conditions contains all the ingredients needed to construct a rational Hodge structure \cite{Jockers:2025fgv}. On the other hand, topological defects in the theory that preserve these boundary-state conditions can be regarded as the objects playing the role of Hodge tensors in the geometric setting; their presence then identifies the corresponding CFT as a Hodge locus. When the topological defects in a given model are sufficient to generate a (product of) complex multiplication fields, we are also able to extend the CFT analogue of the notion of complex multiplication.

We conclude with some final remarks on the results and some open directions for future work.
\begin{itemize}

    \item The picture suggested by our analysis closely parallels what happens in the geometric setting, although the exact connection is not known. There, special points in moduli space where Hodge tensors arise from symmetries are typically accompanied by a reduction in the complexity of the period data \cite{grimm:2024tgd}, which should translate into simplifications of the associated couplings in the effective action. This motivates the expectation that an analogous phenomenon should occur for the CFT Hodge loci introduced in this work. Indeed, since the appearance of Hodge endomorphisms in the worldsheet theory imposes additional algebraic structure on the period matrix, one may expect the corresponding loci to define special subspaces of the moduli space where the low-energy data become more constrained and potentially more tractable.  From this point of view, the role of topological defects is not merely the one to provide a distinguished algebraic structure inside the CFT, but also to signal special regions in moduli space where the effective theory could exhibit simplification. It is therefore natural to expect that a similar mechanism, to the one responsible for the simplification of periods in the geometric case, should also operate here, leading to non-trivial restrictions on the couplings of the effective action. Exploring this expectation in detail would be very interesting, as it could clarify how far the geometric intuition behind Hodge loci extends to the CFT setting and whether the corresponding algebraic constraints admit a direct physical interpretation. More broadly, it would be worthwhile to investigate the relation between the CFT Hodge loci and their geometric counterparts in greater depth. In particular, since in the geometric counterpart in correspondence of Hodge loci have been observed specific geometries, it would be desirable to understand whether the presence of topological defects (Hodge endomorphisms) in the category $\TDL$ of the SCFT can be used not only as an intrinsic characterization of special points in the $\mathcal{N}=(2,2)$ moduli space, but also as a probe of the underlying target-space geometry.

     \item One of the key ingredients underlying this work are boundary states satisfying A-type and B-type gluing conditions, and in particular the fact that these states furnish a module over the category $\TDL$ of topological defects, which we used to define the notions of Hodge loci and complex multiplication. Roughly speaking, this means that the category $\TDL$ is able to impose constrains on the set of boundary states. In particular, at complex multiplication points, these constraints become sufficiently restrictive to generate the entire lattice of boundary states starting from a small set of generators: one for each irreducible component of the rational Hodge structure, or equivalently one for each complex multiplication field appearing in $\prod_{\alpha} K_{\alpha}$. Therefore, the presence of non-trivial defects in $\TDL$ gives rise to new methods for constructing boundary states. Notably, the category $\TDL$ is non-trivial for certain non-geometric models, such as certain asymmetric orbifolds, for which the construction of boundary states remain an open problem. We plan to describe these methods of construction in detail in \cite{Angius:2026xxx}.   
    
    \item  
    The existence of generalized symmetries in a SCFT $\calC$ usually put constraints on the OPE of local operators or, more generally, relate them to correlation functions of defect-starting point operators (for non-invertible ones). The category $\TDL$ acts invertibly on the subCFT generated by the $\CN=(2,2)$ superconformal algebra and the NS-NS and R-R spectral flow operators. This subCFT contains, in particular, the exactly marginal operators inducing deformations of the SCFT.  
    The information about the Hodge endomorphism $\EndL$ is sufficient to determine the action of $\mathcal{L}$ on the exactly marginal operators, and therefore to understand which deformations of the models preserve the corresponding generalized symmetry and which ones break it. Suppose, for example, that one is able to determine the existence of some topological defect at some special point in the moduli space (like at a rational conformal field theory, such as a Gepner model \cite{Cordova:2025crg,Angius:2025zlm}, or a LG model \cite{Brunner:2007qu,Brunner:2013ota}, or some limit where the sigma model is perturbatively under control \cite{Arias-Tamargo:2025xdd,Arias-Tamargo:2025fhv}). Then, one could use these arguments about  Hodge loci to establish the existence of such generalized symmetries even far away of these special points, where we have no tools to study the SCFT. Vice versa, one could use properties of the Hodge structure at generic points in the moduli space to exclude the existence of generalized symmetries at these points. This line of arguments was used in \cite{Angius:2024evd} to show that the set of K3 sigma models admitting a non-trivial defect that preserves the $\CN=(4,4)$ algebra and spectral flow has measure zero (but is possibly dense) in the moduli space. See also \cite{Cordova:2025crg} for a similar approach.
    
    \item The algebra $\End_{\TDL}$ of endomorphisms induced by topological defects in $\TDL$ embeds $\End_{\TDL} \hookrightarrow \End_{Hdg}$ in the full algebra of Hodge endomorphisms $\End_{Hdg}$ on $\Hh^\bullet(\calC,\QQ)$. It is an important open question if and when such embedding is a surjection. For classes of sigma models for which it is surjective, i.e. when topological defect lines are sufficient to induce a complete set of generators for the algebra of Hodge endomorphisms, then the SCFT $\calC$ is CM, according to our definition, if and only if the corresponding $\Hh^\bullet(\calC,\QQ)$ is a CM Hodge structure in the standard sense. When the embedding is not surjective, then the CM property for the SCFT $\calC$ implies the CM property for the Hodge structure, but the converse might not be true.\\
    It is suggestive to reformulate this problem of surjectivity using the `doubling trick'. It is well-known that each topological defect line in $\calC$ can be described as a boundary state in the product $\calC \times \bar\calC$, where  $\bar\calC$ is obtained from $\calC$ by reversing the worldsheet orientation\footnote{Note, however, that the converse is not true: a boundary state in $\calC \times \bar\calC$ in general corresponds to a conformal defect in $\calC$, with reflection coefficient varying from $0$ (topological defect) to $1$ (a tensor product of two boundary states). }. The Hodge endomorphism associated with the topological defect $\CL$ in $\calC$ is just the R-R charge of the corresponding boundary state in $\calC\times\bar\calC$. This is the CFT analogue of the statement that a Hodge endomorphism of a Hodge structure $(V_\QQ,V_\CC)$ can be equivalently described as a rational Hodge tensor, i.e. a rational Hodge class in the $(0,0)$ Hodge component of the tensor product $\mathcal{T}_1^1V=V_\QQ\otimes V^\vee_\QQ$ (see appendix \ref{s:appendix_A}). \\
    Thus, the problem of surjectivity of the embedding above is closely related to the question of whether the full rational space $\Hh^\bullet (\calC\times \bar\calC,\QQ)$ is generated over $\QQ$ by the R-R charges of supersymmetric D-branes in $\calC\times \bar\calC$. One can also understand this problem as a SCFT version of the (rational) Hodge conjecture, stating that for a smooth projective variety $X$ over $\CC$, all spaces $H^{p,p}(X)\cap H^{2p}(X,\QQ)$ are generated by the classes corresponding to cycles of codimension $p$ in $X$.
    
    \item By construction, the topological defects in $\TDL$ are unaffected by both the $A$-type or $B$-type topological twist of the $\CN=(2,2)$ SCFT $\calC$, and therefore they give rise to topological defect lines in the corresponding topological model. Such defects have been recently described in \cite{Caldararu:2025eoj}, see also \cite{Brunner:2013ota}.  It is an interesting open question whether all generalized symmetries of the topological $A$- or $B$-model can be obtained in this way. \\
    Recall that the topological model provide the framework of a rigorous mathematical definition of mirror symmetry. In this context, $B$-type D-branes are interpreted as objects in the (bounded) derived category $D^b(Coh(X))$ of coherent sheaves on the target space $X$, while $A$-type D-branes belong to the Fukaya category \cite{Sharpe:2003dr,Aspinwall:2004jr}. The homological mirror symmetry conjecture is formulated as an equivalence between the Fukaya category of $X$ and $D^b(Coh(Y))$, where $Y$ is the mirror manifold of $X$. This idea suggests that both  $D^b(Coh(X))$ and the Fukaya category on $X$ could be described as module categories for the same tensor category $\TDL$. This might shed some new light in our understanding of homological mirror symmetry \cite{Kontsevich:1994dn}. 
    
   \item It was conjectured in \cite{Gukov:2002nw} that the rationality of a sigma model might be related to both the target space and its mirror being CM manifolds with the same CM field. This proposal was mainly based on observations about elliptic curves (see section \ref{s:elliptic_curves}). While the discovery of a few counterexamples showed that the conjecture cannot be true if taken literally \cite{Chen:2005gm}, it is still an open question whether some refinement might work at least in some class of higher dimensional CYs (see \cite{Okada:2022jnq,Kidambi:2022wvh,Kidambi:2024vwl} for some recent attempts in this direction).\\
   In \cite{Jockers:2025fgv}, it was proposed that rationality of $\CN=(2,2)$ CFTs might be related to CM properties of the Hodge structure of the space of R-R ground fields. The connection of CM properties of the SCFT with the presence of topological defects might help to make this relation precise.
   Indeed, on the one hand, if a SCFT $\calC$ is rational, then it should admit a fusion category (i.e. a tensor category of finitely many objects) of topological defects preserving the product $\CA\times\bar \CA$ of  the chiral and antichiral algebra, or acting on it by automorphism. For example, when $\CA$ and $\bar\CA$ are isomorphic rational VOAs, and $\calC$ is the diagonal modular invariant, then the Verlinde lines form a fusion category isomorphic to the modular tensor category of $\CA$ modules. It is clear that every defect $\CL$ preserving $\CA\times\bar\CA$ must be contained in the category $\TDL$, because it acts trivially on the $\CN=(2,2)$ algebra and on the holomorphic spectral flow operators. Furthermore, one can usually prove that the induced Hodge endomorphisms $\EndL$ are non-trivial: indeed, a trivial action of $\CL$ on all R-R ground fields implies that $\CL$ preserves also all exactly marginal operators, and therefore that the generalized symmetry cannot be broken by any deformation of the model. This line of reasoning suggests that only Hodge loci in the moduli space of sigma models can be rational points. In the other direction, one could try to characterise rationality of a certain SCFT $\calC$ by the presence of a fusion subcategory (in particular, with finite many simple objects) of $\TDL$, whose preserved subCFT\footnote{We recall that, in this article, when we talk about `subCFT' we do not require modular invariance on the torus or higher genus Riemann surfaces, but only that the OPE on the sphere closes.} factorizes holomorphically as $\CA\times \bar\CA$. In this case, on general grounds, one expects (at least at a physical level of rigor) the category of representations of $\CA\times \bar\CA$ to be also finite. We stress that we do not have, at the moment, a way to establish the holomorphic factorization condition for the preserved subCFT of a given fusion category; a possible approach to this problem could involve factorization of the associated twined partition functions.
   \item Let $\calC$ be the internal SCFT obtained from the compactification of type IIA superstring on a K3 manidolfd. Then, the presence of a non-trivial Hodge class in the Picard lattice $\Hh^\bullet (\calC,\ZZ)\cap \Hh^{1,1}(\calC,\CC)$ has interesting consequences for the dual heterotic string on $T^4$. Indeed, by matching the BPS charges on the two sides of the duality, it is easy to see that such a D-brane charge in type IIA corresponds to a purely anti-holomorphic winding-momentum for the heterotic string (we use conventions where the bosonic sector of the heterotic string is anti-holomorphic). This implies that, at points of the moduli space where a Hodge class for $\calC$ exists, the generic anti-chiral algebra of the heterotic string gets extended by a purely anti-holomorphic vertex operator.\footnote{If we require $\calC$ on the type IIA side to be a well-defined compact unitary CFT, then this antiholomorphic vertex operator must have conformal weight strictly greater than $1$, so that it does not give rise to additional massless fields in space-time. In particular, the gauge group of the low energy effective theory is the generic $U(1)^{24}$.} In particular, when the Picard lattice $\Hh^\bullet (\calC,\ZZ)\cap \Hh^{1,1}(\calC,\CC)$ has maximal dimension $20$, then the heterotic worldsheet CFT becomes rational. Some examples of these models include the heterotic models with `Conway groups of symmetries', considered in \cite{Harvey:2017xdt,Baykara:2021ger}. \\
   A similar story occurs when $\calC$ in the internal SCFT for the compactification of type II on $T^4$. The S-dual of this theory is still type II on a different torus $T^4$, where the Narain lattice of winding momentum and the lattice of even-dimensional D-brane charges get exchanged (this make sense, since in this case they are both even unimodular lattices of signature $(4,4)$).  Again, the presence of a non-trivial Hodge class in $\Hh^\bullet (\calC,\ZZ)\cap \Hh^{1,1}(\calC,\CC)$ corresponds to a purely anti-holomorphic winding-momentum in the S-dual theory. Furthermore, $\Hh^\bullet (\calC,\ZZ)\cap \Hh^{1,1}(\calC,\CC)$ has maximal dimension $4$ if and only if the S-dual theory corresponds to a rational SCFT. The $T^4$ sigma models with non-geometric symmetries in section 4.4 of \cite{Volpato:2014zla} are all examples of this phenomenon. The  S-dual model contains purely holomorphic bosonic vertex operators, with winding-momenta in a root lattice $\Lambda$ of rank $4$, so that the chiral algebra contains a conformally embedded rational Kac-Moody algebra at level $1$.\\ These connections between the presence of Hodge classes and extended chiral algebras (or even rationality) for the S-dual worldsheet CFT provide further support to the general interpretation of Hodge loci as points in the moduli space where symmetries get enhanced and special simplifications occur.
\end{itemize}

\bigskip

{\bf Acknowledgments.}  
We are pleased to thank Angel Uranga for useful discussions and comments on the manuscript, Stefano Giaccari, Sarah Harrison and Alessandro Miccich\`{e} for collaboration on closely related projects.   \\
R.A. acknowledges support from the Deutsche Forschungsgemeinschaft under Germany’s Excellence Strategy EXC 2121 Quantum Universe 390833306  and Deutsche Forschungsgemeinschaft through the Collaborative Research Center 1624 "Higher Structures, Moduli Spaces and Integrability". R.V. acknowledges support from CARIPARO Foundation Grant under grant n. 68079.

\appendix

\section{Geometric background}
\label{s:appendix_A}

In this appendix  we introduce the notion of Hodge loci in the complex structure moduli space for families of $CY$ manifolds using a geometric language.

Given a family of Calabi-Yau D-folds, the associated complex structure moduli space $\mathcal{M}_{cs}$ is a K\"ahler manifold of complex dimension $h^{D-1,1}$ which is in general neither smooth nor compact. This implies the existence of singular loci, whose corresponding Calabi–Yau are singular manifolds. Let us denote this set of loci by $\Delta$ and refer to it as the \textit{discriminant locus}. The coordinates of $\mathcal{M}_{cs}$ parametrize deformations of the complex structure and therefore control the geometry of the middle-dimensional cycles, equivalently the structure of the middle homology. Moreover, this moduli space is accompanied by an underlying algebraic structure that varies with changes of the moduli. A way to provide a mathematical formulation of this structure is in terms of the Hodge structure of the middle cohomology and its variation. In particular, given a smooth Calabi-Yau manifold $Y_D$, the $k-th$ cohomology groups $H^k \left( Y_D, \mathbb{C} \right)$ of the manifold always admit a \textit{pure Hodge structure} of weight $k$, which means that for each level $k$ we can define a vector space $V_{\mathbb{C}}= H^k  \left( Y_D, \mathbb{C} \right)$, which always admits an Hodge decomposition:
  \begin{equation}
  V_{\mathbb{C}}^{(k)} = H^{k,0} \oplus H^{k-1,1} \oplus ... \oplus H^{1,k-1} \oplus H^{0,k} = \bigoplus_{k=p+q} H^{p,q} 
  \label{Hodge_decomposition}
  \end{equation}
  where the building subspaces satisfy the following complex-conjugation property:
  \begin{equation}
  H^{p,q} = \overline{H}^{q,p},
  \end{equation}
  and the weight $k$ is the constant sum of the indices $p+q=k$ of all the blocks in \eqref{Hodge_decomposition}.
  
Inside the realm of string compactifications, the main interest is in general addressed to analyze the Hodge structure associated with the middle cohomology group as the corresponding elements encodes the main ingredients to construct the low energy effective field theory arising from these compactifications: 
 \begin{equation}
      V_{\mathbb{C}}=H^D \left( Y_D, \mathbb{C}\right) = \bigoplus_{D=p+q} H^{p,q} 
      \label{e:Hodge_decomposition_middle}
  \end{equation}

    An alternative way to rephrase the same information is by saying that each cohomology group $H^D \left( Y_D, \mathbb{C} \right)$ defines a decreasing Hodge filtration:
   \begin{equation}
0 \subset H^{D,0}=F^D \subset F^{D-1} \subset ... \subset F^1 \subset   F^0 = V_{\mathbb{C}} 
\label{e:filtration}
\end{equation}
of vector spaces 
\begin{equation}
    F^p = \bigoplus_{r \geq p} H^{r,D-r}.
\end{equation}
This formulation is equivalent to the previous one via $H^{p,q} = F^p \cap \overline{F}^q$.\\
If we consider a full family of Calabi–Yau manifolds associated with a given moduli space $\mathcal{M}_{cs}$, then as we move across the moduli space, the Hodge decomposition \eqref{e:Hodge_decomposition_middle}, as well as the $F^p$ spaces in the filtration \eqref{e:filtration}, vary with the moduli. This means that, while the full vector space $V_{\mathbb{C}}$ remains fixed, the orientation of the individual $H^{p,q}$ subspaces depends on the moduli. This changing process corresponds to what is known under the name of variation of the Hodge structure.

If one can further equip the Hodge structure with a bilinear form $Q(\cdot , \cdot)$ on $V_{\mathbb{C}}$ satisfying the following two properties (see eq.\eqref{e:polarization_properties}):
\begin{equation}
\begin{split}
& (i) \quad \textsc{orthogonality} \quad Q \left( H^{p,q} , H^{r,s} \right)=0 \quad \text{for} \quad p \neq s, q \neq r ; \\
& (ii) \quad \textsc{non-degeneracy} \quad i^{p-q} Q \left( v, \overline{v} \right) >0 \quad \text{for} \quad v \in H^{p,q}, \quad v \neq 0 ,   \\
\end{split}
\end{equation}
the structure $H^D$ is called \textit{polarized}.
For smooth Calabi-Yau manifolds such a the bilinear form is provided by the cup product:
\begin{equation}
    Q(v,w)= \int_{Y_D} v \wedge w \quad \quad \quad v,w \in H^D \left(Y_D, \mathbb{C} \right),
    \label{e:polarization_cup}
\end{equation}
which induces a norm for the vectors of the whole filtration given by:
\begin{equation}
\vert \vert v \vert \vert^2 = Q(v,  \ast \overline{v})
\label{def:Hodge_norm}
\end{equation}
and called Hodge norm. The isometry group preserving the pairing \eqref{e:polarization_cup} is isomorphic to $SO(m,n)$, for certain integer numbers $m$ and $n$ such that $m+n=\dim H^D$, for $D$ even and $Sp(\dim H^D)$ for $D$ odd. In particular, we have:
\begin{itemize}
    \item $SO(3,19)$ for $K3$; 
    \item $Sp(2h^{2,1}+2)$ for $CY_3$;
    \item $SO(h^{2,2}+2,2h^{3,1})$ for $CY_4$.
\end{itemize}

An alternative way to express the variation of the Hodge structure is in terms of the variation of the holomorphic $D$-form $\Omega$ with respect to a fixed \textit{integral} basis $\left\lbrace \gamma^{\mathcal{I}} \right\rbrace $ of $F^0$, with $\mathcal{I} = 1,..., dim H^D $, and internal pairing represented by the matrix:
\begin{equation}
    \eta_{\mathcal{I} \mathcal{J}} = - \int_{Y_D} \gamma^{\mathcal{I}} \wedge \gamma^{\mathcal{J}}.
\end{equation}
We can now use this basis to expand the form $\Omega$ as:
\begin{equation}
\Omega = \Pi^{\mathcal{I}} (\mathbf{z}) \gamma_{\mathcal{I}}.
\label{periods}
\end{equation}
The coefficients of such expansion are called \textit{periods} and they are in general complicated transcendental functions of the moduli, that here we are indicating with $\mathbf{z}$. Equivalently, they are defined by:
 \begin{equation}
 \Pi^{\mathcal{I}} (z) = \int_{\Gamma^{\mathcal{I}}} \Omega,
 \label{periods_def}
\end{equation}  
where $\Gamma^{\mathcal{I}}$ are the 4-cycles Poincaré-dual of the forms $\gamma^{\mathcal{I}}$. At this point one can understand the variations of the spaces $F^p$ on $V_{\mathbb{C}}$ over the moduli space $\mathcal{M}_{cs}$ in terms of the variations of the periods \eqref{periods_def} with respect to the coordinates $\mathbf{z}$.

This whole structure is completely well defined in the bulk of the moduli space, as long as the manifolds $Y_D$ are smooth. However, when we approach a singular locus in $\mathcal{M}_{cs}$ the periods \eqref{periods_def} diverge, the Hodge filtration $F^p$ and the corresponding Hodge decomposition $H^{p,q}$ are not well defined anymore and we need to consider a refinement of the structure. This refinement involves a Mixed Hodge structure, where all the degree cohomologies are considered at the same time.  We will skip the details of the discussion of these special loci and we refer the reader to the  mathematical literature \cite{Schmid:1973,Cattani:1986}. Before proceeding, let us briefly comment on the behaviour of periods near singular loci. Periods are multivalued functions of the moduli that around singular loci transform under specific monodromy matrices $M$ depending on the type of degeneration occurring at the corresponding locus. These local monodromies lie in the isometry group and impose strong constraints on the local structure of the periods, leading to controlled asymptotic expansions. In particular, the periods admit local expansions near such loci whose leading behaviour is polynomial, while deviations from this behaviour are encoded in infinite towers of exponentially suppressed corrections as one moves away from the singular point into the bulk of the moduli space. The key point is that monodromy provides algebraic constraints on the periods, allowing for a local approximation in terms of finitely generated structures, corrected by infinite series of exponential corrections which are the mark that, at generic points of the moduli space, periods are highly transcendental functions.

Let us now consider the rational Hodge structure $(V_{\mathbb{Q}}, V_{\mathbb{C}})$, where $V_{\mathbb{C}}$ is for example the vector space defined in \eqref{e:Hodge_decomposition_middle}, and let us introduce the notions of Hodge classes and Hodge tensors. We will use these notions to define the Hodge loci. 

\vspace{0.15cm}
\noindent
\textbf{Definition:} {[Integral/Rational Hodge class]} \\
\textit{Given an integral or rational Hodge structure $\left( V_{\mathbb{Z}/\mathbb{Q}}, V_{\mathbb{C}} \right)$ of weight $w$, we define integral and rational Hodge classes of the structures to be elements in the intersections
\begin{equation}
\begin{split}\omega \, \in \, V^{p,p} \cap V_{\mathbb{Z}} \quad & \quad \quad \text{integral Hodge class} \\
\omega \, \in \, V^{p,p} \cap V_{\mathbb{Q}} \quad & \quad \quad \text{rational Hodge class} \\
\end{split}
\end{equation}
with $p = w/2$. } 
\vspace{0.1cm}

Note that if an element $\omega \in V^{p,p}$ defines a rational/integral Hodge class depends on the moduli as the $(p,q)$ splitting varies across the moduli space.  This means that a given rational/integral class in $V_{\mathbb{Q} / \mathbb{Z}}$ becomes an Hodge class only on some sub-loci of the moduli space where there is an alignment between $V^{p,p}$ and the lattice. 

Let us now define the space $\mathcal{T}^m_n V$ of $({m \atop n})$ tensors on $V_{\mathbb{C}}$ as
\begin{equation}
    \mathcal{T}^m_n V = V_{\mathbb{C}}^{\otimes m} \otimes (V_{\mathbb{C}}^{\vee})^{\otimes n}
\end{equation}
where $V_{\mathbb{C}}^{\vee}$ is the dual vector space to $V_{\mathbb{C}}$, i.e.
\begin{equation}
    V_{\mathbb{C}}^{\vee} = Hom \left(  V_{\mathbb{C}} \mapsto \mathbb{C}\right).
\end{equation}
Let us indicate the space of all the tensors on $V_{\mathbb{C}}$  for any value of $m$ and $n$ as
\begin{equation}
    V^{\otimes} = \bigoplus_{m,n} \mathcal{T}^m_n V.
\end{equation}
In an analogous way, the restriction of these definitions to $V_{\mathbb{Z}}$ or  $V_{\mathbb{Q}}$ allows to define the space of integral and rational tensors $\mathcal{T}^m_n V_{\mathbb{Z}}$ and $\mathcal{T}^m_n V_{\mathbb{Q}}$, respectively.\\
The Hodge structure of weight $w$ on $V_{\mathbb{C}}$ induces an Hodge structure of weight $w'=w(m-n)$ on the space of $({m \atop n})$ tensors:
\begin{equation}
    \mathcal{T}^m_n V = \bigoplus_{r+s=w} T^{r,s}
\end{equation}
where $T^{r,s} \equiv \left( \mathcal{T}^m_n V\right)^{r,s}$ contains all the {\footnotesize{$\left( \begin{matrix} m \\ n \end{matrix} \right)$}} tensors $t \in \mathcal{T}^m_n V$ such that
\begin{equation}
    t \in \bigoplus_{p_i, \hat{p}_j \\ q_i, \hat{q}_j} \bigotimes_{i=1}^m V^{p_i,q_i} \otimes \bigotimes_{j=1}^n \left( V^{\hat{p}_j, \hat{q}_j}\right)^{\vee}
\end{equation}
with $\sum_i p_i - \sum_j \hat{p}_j=r$ and $\sum_i q_i - \sum_j \hat{q}_j=s$.\\
With the $T^{r,s}$ blocks in hand, we are now prepared to define integral and rational Hodge tensors.

\vspace{0.15cm}
\noindent
\textbf{Definition:} {[Integral/Rational Hodge tensor]} \\
\textit{The integral/rational Hodge tensors are elements in the intersection
\begin{equation}
\begin{split}
t \, \in \, \left(  \mathcal{T}^m_n V\right)^{p,p} \cap \mathcal{T}^m_n V_{\mathbb{Z}} \quad & \quad \quad \text{integral Hodge tensor} \\
t \, \in \, \left(  \mathcal{T}^m_n V\right)^{p,p} \cap \mathcal{T}^m_n V_{\mathbb{Q}} \quad & \quad \quad \text{rational Hodge tensor} \\
\end{split}
\end{equation}
with $p = D(m-n)/2$. } 
\vspace{0.1cm}

Notice that for $D$ odd, as is the case of the elliptic and $CY_3$, Hodge classes do not exist, however, for $m-n \in 2 \mathbb{Z}$ they can support Hodge tensors.\\

Particularly interesting is the space $\mathcal{T}^1_1 V$ of $({1 \atop 1})$ tensors, which can be interpreted as $\mathbb{C}-$linear maps
\begin{equation}
    t \, \, : \, \, V_{\mathbb{C}} \, \mapsto \,  V_{\mathbb{C}}.
\end{equation}
Since the Hodge decomposition of $\mathcal{T}^1_1 V$ is of weight $0$, then it only contains the space $T^{0,0}= \left( \mathcal{T}^1_1 V \right)^{0,0}$ that acts on $V_{\mathbb{C}}$ preserving the Hodge structure:
\begin{equation}
    \begin{split}
        t \,: \,  \, V_{\mathbb{C}} \quad & \mapsto \quad V_{\mathbb{C}} \\
        V^{p,q} \quad & \mapsto \quad V^{p+r,q+s}
    \end{split}
\end{equation}
with $r=s=0$.\\
When an integral/rational tensors becomes of Hodge type, it induces by $\mathbb{C}$-extension a linear map on $V_{\mathbb{C}}$ that preserves the Hodge structure. The monodromy matrices around singularities in the complex structure moduli space are examples of Hodge tensors. Hodge tensors of type $({1 \atop 1})$ corresponds to Hodge endomorphisms in the algebraic definition of Section \ref{s:proposal}. This means that topological defects in the CFT $\mathcal{C}$ satisfying properties $(1)$ and $(2)$ formulated in Section \ref{s:proposal} are examples of Hodge tensors for the rational Hodge structure $\left( V_{\mathbb{C}}, V_{\mathbb{Q}}\right)$, where $V_{\mathbb{C}}= H(\mathcal{C}, \mathbb{C})$ as in \eqref{e:H(C,C)} and $V_{\mathbb{Q}}$ is the rational extension of the lattice of BPS boundary states of the theory.\\
The set of integral/rational Hodge tensors of type $({1 \atop 1})$ equipped with composition define the algebra
\begin{equation}
    Hg^{1,1} =(\mathcal{T}^1_1 V)^{0,0} \cap \mathcal{T}^1_1 V_{\mathbb{Q}}.
\end{equation}
This corresponds to the algebra of endomorphisms $End_{Hdg} (V_{\mathbb{Q}})$ defined in Section \ref{s:proposal}.

With the two notions of Hodge integral/rational classes and integral/rational Hodge tensors in mind, we define the \textit{Hodge loci} as the subspaces $HL (\mathcal{M}_{cs}, V_{\mathbb{C}}) \subset \mathcal{M}_{cs}$ where non-trivial integral/rational Hodge classes appear, and $HL (\mathcal{M}_{cs}, V^{\otimes}) \subset \mathcal{M}_{cs}$ where non-trivial integral/rational Hodge tensors arise. Here, “non-trivial” means classes or tensors that are not present at a generic point of the moduli space, but instead appear only on special loci.

\subsection{The Deligne torus and the Mumford-Tate group}

Let us now discuss an alternative group-theoretic formulation of the notion of Hodge structure, which is closely related to the techniques employed in Section \ref{s:proposal} to endow the complex vector space of R-R ground states of the CFT model $\mathcal{C}$ with a Hodge structure.  Let us call $I$ the isometry group of the bilinear pairing in the middle cohomology $H^D (Y_D, \mathbb{C})$. The key object for the construction is the Deligne torus:
\begin{equation}
    \mathbb{S}(\mathbb{R})= \mathbb{C}^{\times}
\end{equation}
namely the group obtained by viewing the complex multiplicative group $\mathbb{C}^{\times}$ as a real algebraic group, whose maximal compact subgroup is
$U(1)$. Using this compact group, one can introduce the algebraic representation
\begin{equation}
    h \, : \, U(1) \, \mapsto \, \, I
\end{equation}
whose corresponding action on $H^D (Y_D, \mathbb{C})$ determines the Hodge decomposition. More precisely, we have
\begin{equation}
    \forall \, \, \omega_{p,q} \in H^{p,q} \quad \rightarrow \quad h(z) \omega_{p,q} = z^p \bar{z}^q \omega_{p,q}
\end{equation}
with $z \in \mathbb{C}^{\times} \, \vert \, \vert z\vert =1$. \\
The Hodge decomposition corresponds now to the decomposition on eigenspaces of $h(z)$. In our CFT analysis the role of this $U(1)$ is played by the $U(1) \times U(1)$ R-symmetries. Morally, the presence of two $U(1)$ groups, allows us to distinguish the gradings of cohomologies of different degrees, the horizontal and the vertical ones.\\
Now, let us consider the orbit $h(U(1))$ on $I$ which defines a 1-parameter subgroup of the isometry group.  We define the Mumford-Tate group $MT(h)$, associated with $h$ as follows. \\

\noindent
\textbf{Definition} [Mumford-Tate group]\\
\textit{The Mumford-Tate group $MT(h)$ is the smallest $\mathbb{Q}$-algebraic subgroup of $I$ containing the orbit $h(U(1))$.}

At a generic point of the moduli space the smallest $\mathbb{Q}$-algebraic subgroup of $I$ containing $h(U(1))$ is $I$ itself. However, at special points of the moduli space, there exists smaller $\mathbb{Q}$-algebraic subgroups of $I$ that contains these orbits. In particular, at CM points the Mumford-Tate group becomes Abelian.
\begin{equation}
    MT(h) \, = \quad \begin{cases}
        & I \quad \text{generically} \\ & \subset I \quad \text{special points}
    \end{cases}
\end{equation}

An alternative definition for the $MT$ group is given in terms of subgroups of the group $GL(H)$ that preserve rational Hodge tensors of weight $0$. See reference \cite{Carlson:2017} for a detailed discussion.

\section{Some technical calculations}\label{a:technical}

In this appendix, we collect some technical calculation that are used in section \ref{s:elliptic_curves}.

\subsection{D-brane charges}\label{a:Dbranes}

Let us consider the RR component of supersymmetric boundary states in the sigma model with target the elliptic curve, described in section \ref{s:elliptic_curves}. The supersymmetric A-type D-branes wrap special Lagrangian submanifolds of $T^2$, and are therefore $D1$-branes, while the B-type wrap holomorphic cycles, that in this case can be only points ($D0$-branes) or the whole $T^2$ ($D2$-branes), or bound states thereof. We only need to find a set of four BPS boundary states which charges span the whole lattice.

The boundary states are written as superpositions of Ishibashi states with definite winding-momentum $p,\tilde p$. We denote by
\be |p,\tilde p,s,\tilde s\rrangle\ ,\ee the R-R Ishibashi states with momenta $(p,\tilde p)$, where $s,\tilde s\in \{\pm \frac{1}{2}\}$ are the $J_0$ and $\tilde J_0$ $U(1)$ charges of the ground state components.  Using
\be G_0^+ |p,\tilde p,s,\tilde s\rrangle=i\sqrt{2}p^*\psi_0^+|p,\tilde p,s,\tilde s\rrangle\qquad G_0^- |p,\tilde p,s,\tilde s\rrangle=i\sqrt{2}p\psi_0^-|p,\tilde p,s,\tilde s\rrangle
\ee
\be \tilde G_0^+ |p,\tilde p,s,\tilde s\rrangle=i\sqrt{2}\tilde p^*\tilde \psi_0^+|p,\tilde p,s,\tilde s\rrangle\qquad \tilde G_0^- |p,\tilde p,s,\tilde s\rrangle=i\sqrt{2}\tilde p\tilde\psi_0^-|p,\tilde p,s,\tilde s\rrangle
\ee
we get the following gluing conditions for the zero modes
\be\label{constrA} \begin{cases}
(J_0-\tilde J_0)|p,\tilde p,s,\tilde s\rrangle=0 \\
    (p^* \psi_0^++i\epsilon\tilde p\tilde\psi_0^-)|p,\tilde p,s,\tilde s\rrangle=0\\
    (p \psi_0^-+i\epsilon\tilde p^*\tilde\psi_0^+)|p,\tilde p,s,\tilde s\rrangle=0
\end{cases}\qquad \text{A-type}
\ee
\be\label{constrB} \begin{cases}
(J_0+\tilde J_0)|p,\tilde p,s,\tilde s\rrangle=0 \\
    (p^* \psi_0^++i\epsilon\tilde p^*\tilde\psi_0^+)|p,\tilde p,s,\tilde s\rrangle=0\\
    (p \psi_0^-+i\epsilon\tilde p\tilde\psi_0^-)|p,\tilde p,s,\tilde s\rrangle=0
\end{cases}\qquad \text{B-type}
\ee From the gluing conditions for $J_0$ and $\tilde J_0$, we have that only the states with $s=\tilde s$ contribute to the A-type branes and only states with $\tilde s=-s$ contribute to B-type. 

For $D0$-branes, the only winding-momentum states that contribute are the $|n_k,w_k\rangle$ with $w_1=0=w_2$ and arbitrary $n_1,n_2\in \ZZ$, and the RR Ishibashi states satisfying the gluing conditions above are
\be |D0,n_1,n_2\rrangle_{\epsilon,RR}:=\frac{1}{\sqrt{2}}\left(|p,\tilde p,+-\rrangle +i\epsilon |p,\tilde p,-+\rrangle\right)\ ,\ee where $(p,\tilde p)$ are related to $n_1,n_2,\in \ZZ$ by \eqref{windmom} for $w_1,w_2=0$.
Similarly, $D2$-branes only get contributions from Ishibashi states with $n_1,n_2=0$ and
\be |D2,w_1,w_2\rrangle_\epsilon=\frac{1}{\sqrt{2}|\rho|}(\rho|p,\tilde p,+-\rrangle +i\epsilon\bar\rho |p,\tilde p,-+\rrangle)\ .\ee
As for A-type branes, the Ishibashi states for $D1$-branes wrapping the fundamental $1$-cycles $\alpha,\beta$ are given by
\be |D1(\alpha),n_2,w_1\rrangle_\epsilon=\frac{1}{\sqrt{2}}(|p,\tilde p,++\rrangle -i\epsilon |p,\tilde p,--\rrangle)\ee and 
\be |D1(\beta),n_1,w_2\rrangle_\epsilon=\frac{1}{\sqrt{2}|\tau|}(\bar\tau|p,\tilde p,++\rrangle -i\epsilon\tau |p,\tilde p,--\rrangle)\ .\ee
The full R-R boundary states are given by (suitably normalized) superpositions of the corresponding Ishibashi states
\be ||D0,a_1,a_2\rrangle_{\epsilon,RR}=\frac{1}{\sqrt{2\rho_2}}\sum_{n_1,n_2\in \ZZ} e^{2\pi i (a_1n_1+a_2n_2)}|D0,n_1,n_2\rrangle_{\epsilon,RR}
\ee 
\be ||D2,b_1,b_2\rrangle_{_\epsilon,RR}=\frac{|\rho|}{\sqrt{2\rho_2}}\sum_{w_1,w_2\in \ZZ} e^{2\pi i (b_1w_1+b_2w_2)}|D2,w_1,w_2\rrangle_{\epsilon,RR}
\ee
\be ||D1(\alpha),b_1,a_2\rrangle_{\epsilon,RR}=\frac{1}{\sqrt{2\tau_2}}\sum_{w_1,n_2\in \ZZ} e^{2\pi i (b_1w_1+a_2n_2)}|D1(\alpha),n_2,w_1\rrangle_{\epsilon,RR}
\ee
\be ||D1(\beta),a_1,b_2\rrangle_{\epsilon,RR}=\frac{|\tau|}{\sqrt{2\tau_2}}\sum_{n_1,w_2\in \ZZ} e^{2\pi i (a_1n_1+b_2w_2)}|D1(\beta),n_1,w_2\rrangle_{\epsilon,RR}
\ee
where $a_1,a_2,b_1,b_2\in \RR/\ZZ$ are the moduli.

\subsection{Field elements on the unit circle}\label{a:units}

Suppose there is $\gamma_\tau:=\left(\begin{smallmatrix}
        a & b\\ c & d
    \end{smallmatrix}\right)\in SL(2,\QQ)$ with $c\neq 0$ such that $\tau=\frac{a\tau+b}{c\tau+d}$ (if $c=0$, then the only possibility is that $\gamma_\tau$ is $\pm 1$). This implies that $\tau$ is a solution of the quadratic equation
    \be cx^2-(a-d)x-b=0\ ,
    \ee with $c\neq 0$, and the other solution is necessarily its complex conjugate $\bar\tau$. It follows that
    \be \frac{a-d}{c}=\tau+\bar\tau=2\Re\tau\ ,\qquad \frac{-b}{c}= \tau\bar\tau=|\tau|^2\ .
    \ee Therefore, $\gamma_\tau$ is of the form
    \be\label{geneq} \gamma_\tau=\begin{pmatrix}
        d+2c\Re\tau & -c|\tau|^2\\ c & d
    \end{pmatrix}
    \ee for suitable $c ,d\in \QQ$. The condition that $\det\gamma_\tau=1$ is equivalent to 
    \be d^2+2cd\Re\tau+c^2|\tau|^2=1\qquad \Leftrightarrow\qquad |c\tau+d|^2=1\ .
    \ee Thus, we find that the group elements $\gamma_\tau\in SL(2,\QQ)$ satisfying \eqref{CMconds} are in one-to-one correspondence with elements $c\tau+d$ in the field $\QQ+\tau\QQ$ on the unit circle $\{z\in \CC\mid |z|=1\}$. To show that there are infinitely many points of modulus $1$ on $\QQ(\tau)$, consider two rational numbers $m,n\in \QQ$, and notice that
    \be |m+n\tau|^2=(m+n\tau)(m+n\bar \tau)=(m+n\tau)(m+n\frac{a-d}{c}-n\tau)
    \ee is a rational number. Indeed, it is an element of $\QQ+\tau\QQ$ and is fixed under complex conjugation, and therefore it is in $\QQ+\tau\QQ\cap \RR=\QQ$. Thus, by setting
    \be c\tau+d:=\frac{(m+n\tau)^2}{|m+n\tau|^2}\ ,
    \ee one gets an element in $\QQ(\tau)\cap S^1$ for each pair of rationals $(m,n)\neq (0,0)$. Notice that one can always rescale $m,n$ by a common rational number, so that they can be chosen to be coprime integers. This procedure provides infinitely many elements in $\QQ(\tau)\cap S^1$, but in some cases they are not all of them. For example, if $\QQ(\tau)=\QQ(\sqrt{-5})$, then the element $\frac{2+i\sqrt{5}}{3}\in S^1$ cannot be obtained in this way.

\subsection{Quantum dimension of non-invertible defects}\label{a:qdim}

In this appendix, we prove that the quantum dimension $\langle \CL_{\gamma_\tau,\gamma_\rho}\rangle$ is a positive integer and we characterize it. First notice that if a certain pair $(\gamma_\tau,\gamma_\rho)$ obeys \eqref{CMconds}, then also $(\gamma_\tau,1)$ and $(1,\gamma_\rho)$ satisfy them, so that $\CL_{\gamma_\tau,1}$ and $\CL_{1,\gamma_\rho}$ are also topological defects, and
\be \langle \CL_{\gamma_\tau,\gamma_\rho}\rangle=\langle \CL_{\gamma_\tau,1}\rangle\langle \CL_{1,\gamma_\rho}\rangle\ . 
\ee Thus, it is sufficient to prove that  $\langle \CL_{\gamma_\tau,1}\rangle,\langle \CL_{1,\gamma_\rho}\rangle\in \ZZ_{>0}$. In fact we will prove that  $\langle \CL_{\gamma_\tau,1}\rangle$ and $\langle \CL_{1,\gamma_\rho}\rangle$ are the smallest positive integers such that $\langle \CL_{\gamma_\tau,1}\rangle\gamma_\tau$ and $\langle \CL_{1,\gamma_\rho}\rangle\gamma_\rho$ are integral matrices.

Let us focus on $\langle \CL_{\gamma_\tau,1}\rangle$; the argument with $\langle \CL_{1,\gamma_\rho}\rangle$ is completely analogous. First notice that $\CL_{\gamma_\tau,1}^*\CL_{\gamma_\tau,1}$ acts trivially on the holomorphic and antiholomorphic free bosons and fermions $\partial Z$, $\bar\partial Z$, $\psi^\pm$, $\tilde\psi^{\pm}$. Therefore, it must be a superposition of invertible defects in $U(1)^4_{w-m}$. Furthermore, the action on the states $|p,\tilde p\rangle$ is
\be \hat\CL_{\gamma_\tau,1}^* \hat\CL_{\gamma_\tau,1}|p,\tilde p\rangle=\begin{cases}
    \langle \CL_{\gamma_\tau,1}\rangle^2 |p,\tilde p\rangle & \text{if }
(p,\tilde p)\in \Gamma^{2,2}\cap (\Gamma^{2,2}\gamma_\tau)\\
0 & \text{otherwise,}
\end{cases}
\ee where we used that $\langle \hat\CL_{\gamma_\tau,1}^* \hat\CL_{\gamma_\tau,1}\rangle=\langle \CL_{\gamma_\tau,1}\rangle^2$. Thus, $\hat\CL_{\gamma_\tau,1}^* \hat\CL_{\gamma_\tau,1}$ is proportional to a projection operator on the states $|p,\tilde p\rangle$ corresponding to the sublattice $\Gamma^{2,2}\cap (\Gamma^{2,2}\gamma_\tau)$; the correct normalization can be fixed by imposing that $\CL_{\gamma_\tau,1}^*\CL_{\gamma_\tau,1}=\CI+\ldots$, as expected for simple defects. By self-duality of $\Gamma^{2,2}$ and $\Gamma^{2,2}\gamma_\tau$, one can characterize $\Gamma^{2,2}\cap (\Gamma^{2,2}\gamma_\tau)$ as the sublattice of $\Gamma^{2,2}$ that is invariant under the finite abelian subgroup $A_\tau\subset U(1)^4_{w-m}$ given by 
\be A_\tau=(\Gamma^{2,2}\cap \Gamma^{2,2}\gamma_\tau)/\Gamma^{2,2} \subset (\Gamma^{2,2}\otimes \RR)/\Gamma^{2,2}\cong U(1)^4_{w-m}\ .\ee We conclude that
\be \CL_{\gamma_\tau,1}^*\CL_{\gamma_\tau,1}=\sum_{(a_k,b_k)\in A_\tau}\CL_{a_k,b_k}
\ee and therefore
\be \langle \CL_{\gamma_\tau,1}\rangle^2=|A_\tau|\ .
\ee Now, because $\gamma_\tau$ acts separately on $(n_1,n_2)\in \ZZ^2$ and on $(w_1,w_2)\in \ZZ^2$, one has that $A_\tau$ is the direct product $A_\tau=A_\tau^w\times A_\tau^n$ of two isomorphic abelian groups $A_\tau^w$ and $A_\tau^n$, where
\be A_\tau^w=\{\begin{pmatrix}
    w_2 & w_1
\end{pmatrix}\gamma_\tau\mid w_1,w_2\in \ZZ\}/(\ZZ\oplus\ZZ)\subset U(1)_{w_1}\times U(1)_{w_2}
\ee
\be A_\tau^n=\{\begin{pmatrix}
    -n_1 & n_2
\end{pmatrix}\gamma_\tau\mid n_1,n_2\in \ZZ\}/(\ZZ\oplus\ZZ)\subset U(1)_{n_1}\times U(1)_{n_2}\ .
\ee Therefore, 
\be \langle \CL_{\gamma_\tau,1}\rangle=|A_\tau^w|=|A_\tau^n|
\ee is a positive integer. Finally, let us prove that $A_\tau^n$ and $A_\tau^w$ are actually cyclic groups. It is clear that both groups $A_\tau^n$ and $A_\tau^w$ are generated by the columns of the matrix $\tau=\left(\begin{smallmatrix}
    a & b\\ c & d
\end{smallmatrix}\right)$, considered mod $\ZZ$, i.e.
\be A_\tau^n \cong A_\tau^w
\cong \langle \begin{pmatrix} [a]\\ [c]\end{pmatrix},\ \begin{pmatrix} [b]\\ [d] \end{pmatrix}\rangle
\subset (\QQ/\ZZ)^2 \ ,\ee where, for every $x\in \QQ$, we denote by $[x]\in \QQ/\ZZ$ its class mod $\ZZ$.
Now, the condition $ad-bc=1\equiv 0\mod \ZZ$ implies that $\left(\begin{smallmatrix}
    [a] \\ [c]
\end{smallmatrix}\right)$ and $\left(\begin{smallmatrix}
 [b]\\  [d]
\end{smallmatrix}\right)$,  are proportional to each other (mod $\ZZ$)
\be  \begin{pmatrix} [a]\\ [c] \end{pmatrix}=k_1\begin{pmatrix} v_1\\v_2 \end{pmatrix}\mod\ZZ\qquad \begin{pmatrix} [b]\\ [d] \end{pmatrix}=k_2\begin{pmatrix} v_1\\v_2 \end{pmatrix}\mod \ZZ\ ,
\ee for some $k_1,k_2,v_1,v_2\in\QQ$.  If we rescale $\left(\begin{smallmatrix}
 v_1\\  v_2
\end{smallmatrix}\right)$ so that $k_1,k_2$ are coprime integers, we get that
\be A_\tau^n \cong A_\tau^w
\cong \langle \begin{pmatrix} [v_1]\\ [v_2] \end{pmatrix}
\rangle\cong \ZZ_N \ ,\ee
where $N=|A_\tau^n|=|A_\tau^w|=\langle \CL_{\gamma_\tau,1}\rangle$ is the smallest integer such that $N\left(\begin{smallmatrix}
    v_1 \\ v_2
\end{smallmatrix}\right)$ is integral. By construction, this is also the smallest integer for which $N\gamma_\tau$ is integral.

\printbibliography

\end{document}